%% file: paper.tex
\documentclass[sigconf]{acmart}

\input{defs.tex}

\AtBeginDocument{%
  \providecommand\BibTeX{{%
    \normalfont B\kern-0.5em{\scshape i\kern-0.25em b}\kern-0.8em\TeX}}}

\copyrightyear{2022} 
\acmYear{2022} 
\setcopyright{acmlicensed}\acmConference[ISCA '22]{The 49th Annual International Symposium on Computer Architecture}{June 18--22, 2022}{New York, NY, USA}
\acmBooktitle{The 49th Annual International Symposium on Computer Architecture (ISCA '22), June 18--22, 2022, New York, NY, USA}
\acmPrice{15.00}
\acmDOI{10.1145/3470496.3533044}
\acmISBN{978-1-4503-8610-4/22/06}

\begin{document}

\title[Understanding Data Storage and Ingestion for Large-Scale Deep Recommendation Model Training]{Understanding Data Storage and Ingestion for Large-Scale Deep Recommendation Model Training}
\subtitle{Industrial Product}
\subtitlenote{This paper is part of the Industry Track of ISCA 2022's program.}

\author{Mark Zhao\textsuperscript{$\dagger$}, Niket Agarwal\textsuperscript{$\dagger$}, Aarti Basant\textsuperscript{$\dagger$}, Bu\u{g}ra Gedik\textsuperscript{$\dagger$}, Satadru Pan\textsuperscript{$\dagger$},
Mustafa Ozdal\textsuperscript{$\dagger$}, Rakesh Komuravelli\textsuperscript{$\dagger$}, Jerry Pan\textsuperscript{$\dagger$}, Tianshu Bao\textsuperscript{$\dagger$}, Haowei Lu\textsuperscript{$\dagger$}, Sundaram Narayanan\textsuperscript{$\dagger$}, Jack Langman\textsuperscript{$\dagger$}, Kevin Wilfong\textsuperscript{$\dagger$}, Harsha Rastogi\textsuperscript{$\dagger$}, Carole-Jean Wu\textsuperscript{$\dagger$}, Christos Kozyrakis\textsuperscript{$\ddagger$}, Parik Pol\textsuperscript{$\dagger$}}
\affiliation{%
  \institution{\vspace{3mm}
  {\LARGE \textsuperscript{$\dagger$}Meta}, {\LARGE \textsuperscript{$\ddagger$}Stanford University}}
  \country{}
}

\renewcommand{\authors}{Mark Zhao, Niket Agarwal, Aarti Basant, Bu\u{g}ra Gedik, Satadru Pan,
Mustafa Ozdal, Rakesh Komuravelli, Jerry Pan, Tianshu Bao, Haowei Lu, Sundaram Narayanan, Jack Langman, Kevin Wilfong, Harsha Rastogi, Carole-Jean Wu, Christos Kozyrakis, Parik Pol}

\renewcommand{\shortauthors}{M. Zhao, et al.}

\input{abstract}

\ccsdesc[500]{Software and its engineering~Distributed systems organizing principles}
\ccsdesc[300]{Information systems~Database management system engines}
\ccsdesc[500]{Computing methodologies~Machine learning}

\keywords{Machine learning systems, databases, distributed systems, data ingestion, data storage}

\maketitle

\input{introduction}
\input{background}
\input{dsi}
\input{training}

\input{dataset}
\input{preprocessing}
\input{future}

\input{related}
\input{conclusion}

\begin{acks}
We would like to thank the many engineers in the numerous infrastructure and hardware teams that build, support, and maintain the systems and hardware that compose Meta's DSI pipeline.
We also thank Daniel Ford, Dheevatsa Mudigere, Chunqiang Tang, Matei Zaharia, and the anonymous reviewers for their feedback on this paper.
Christos Kozyrakis is supported by the Stanford Platform Lab and its affiliate members.
\end{acks}

\bibliographystyle{ACM-Reference-Format}
\bibliography{paper}

\end{document}

%% file: defs.tex
\usepackage{xcolor}
\usepackage{xspace}
\usepackage{comment}
\usepackage{booktabs}
\usepackage{makecell}
\usepackage{etoolbox}
\usepackage{url}
\usepackage{hyperref}
\usepackage{amsmath,amsfonts}
\usepackage{algorithmic}
\usepackage{graphicx}
\usepackage{textcomp}

\newtoggle{showmarks}
\toggletrue{showmarks}

\newtoggle{reviewdraft}

\newtoggle{cameraready}

\iftoggle{showmarks}{
  \newcommand\markyz[1]{\textcolor{violet}{[~MARK:~#1~]}}
  \newcommand\niketa[1]{\textcolor{blue}{[~NIKET:~#1~]}}
  \newcommand\carole[1]{\textcolor{orange}{[~CAROLE:~#1~]}}
  \newcommand\christos[1]{\textcolor{purple}{[~CHRISTOS:~#1~]}}

  \newcommand\actionitem[1]{\textcolor{red}{[~ACTION:~#1~]}}
}{
  \newcommand\markyz[1]{\unskip}
  \newcommand\niketa[1]{\unskip}
  \newcommand\carole[1]{\unskip}
  \newcommand\christos[1]{\unskip}
  \newcommand\actionitem[1]{\unskip}
}

\iftoggle{reviewdraft}{
  
}{
  
}

\iftoggle{cameraready}{
  
}{
  
}

\newcommand{\SystemName}{\textsc{DPP}\xspace}

%% file: abstract.tex
\begin{abstract}
Datacenter-scale AI training clusters consisting of thousands of domain-specific accelerators (DSA) are used to train increasingly-complex deep learning models.
These clusters rely on a data storage and ingestion (DSI) pipeline, responsible for storing exabytes of training data and serving it at tens of terabytes per second.
As DSAs continue to push training efficiency and throughput, the DSI pipeline is becoming the dominating factor that constrains the overall training performance and capacity.
Innovations that improve the efficiency and performance of DSI systems and hardware are urgent, demanding a deep understanding of DSI characteristics and infrastructure at scale.

This paper presents Meta's end-to-end DSI pipeline, composed of a central data warehouse built on distributed storage and a Data PreProcessing Service that scales to eliminate data stalls.
We characterize how hundreds of models are collaboratively trained across geo-distributed datacenters via diverse and continuous training jobs.
These training jobs read and heavily filter massive and evolving datasets, resulting in popular features and samples used across training jobs.
We measure the intense network, memory, and compute resources required by each training job to preprocess samples during training.
Finally, we synthesize key takeaways based on our production infrastructure characterization.
These include identifying hardware bottlenecks, discussing opportunities for heterogeneous DSI hardware, motivating research in datacenter scheduling and benchmark datasets, and assimilating lessons learned in optimizing DSI infrastructure.
\end{abstract}

%% file: introduction.tex
\section{Introduction}\label{sec:introduction}
Domain-specific accelerators (DSAs) for deep neural networks (DNNs) have become ubiquitous because of their superior performance per watt over traditional general purpose processors~\cite{isca17-tpu}.
Industry has rapidly embraced DSAs for both DNN training and inference.
These DSAs include both traditional technologies, such as GPUs and FPGAs, as well as application-specific integrated circuits (ASICs) from, e.g., Habana~\cite{habana}, Graphcore~\cite{hc33-graphcore}, SambaNova~\cite{hc33-sambanova}, Tenstorrent~\cite{tenstorrent}, Tesla~\cite{tesla}, AWS~\cite{aws-trainium}, Google~\cite{isca17-tpu}, and others.

DSAs are increasingly deployed in immense scale-out systems to train increasingly-complex and computationally-demanding DNNs using massive datasets.
For example, the latest MLPerf Training round (v1.1)~\cite{mlperf-v11} contains submissions from Azure and NVIDIA using 2048 and 4320 A100 GPUs, respectively, whereas Google submitted training results using pods containing up to 4096 TPUv4s~\cite{ieee-tpuv4}.
At Meta, we are building \textit{datacenter-scale AI training clusters} by both scaling our production datacenters to include thousands of GPUs using ZionEX nodes~\cite{arxiv-mudigere} and building the AI Research SuperCluster (RSC) with over ten thousand GPUs~\cite{rsc}.
These DSAs have been laser-focused on optimizing and scaling compute for training, namely matrix-heavy computations used during backpropagation.

\input{fig-ingestion-power}

In reality, DNN training in production involves significantly more than just backpropagation.
Hazelwood \textit{et al.} noted how \textit{``For many machine learning models at [Meta], success is predicated on the availability of extensive, high-quality data''}~\cite{hpca18-hazelwood}. 
Namely, a \textbf{data storage and ingestion} (DSI) pipeline, consisting of \textit{offline data generation, dataset storage, and online preprocessing services}, must store and feed exabytes of data to high-performance training nodes (trainers).
The design of the DSI pipeline significantly affects the overall DNN training capacity and performance, but has received little consideration compared to model training itself.

This paper focuses on {\it understanding DSI requirements, unique workload characteristics, and systems for industry-scale, deep learning recommendation model (DLRM) training}. 
We focus on DLRMs because DLRMs a) underpin many of Meta's personalization and ranking services~\cite{hpca21-acun, hpca18-hazelwood, hpca20-gupta, isca20-deeprecsys, instagram_ranking}, b) consume the vast majority of the overall ML training cycles~\cite{hpca21-acun} (and DSI capacity) in Meta datacenters, and c) introduce novel DSI challenges not yet captured by current ML benchmarks~\cite{mlsys20-mlperf} nor considered by existing systems~\cite{tfdata, cluster19-dlfs, mascots18-deepio, hotcloud19-oneaccess, fast20-quiver, atc21-revamper, vldb21-mohan}.

Understanding and efficiently scaling the DSI pipeline is essential in enabling large-scale training for several reasons.
First, inefficiencies in the pipeline cripple training throughput~\cite{vldb21-mohan}, underutilizing expensive DSAs.
Second, DSI infrastructure competes for valuable power resources with trainers.
Figure~\ref{fig:ingestion_power} shows how \textit{storage and online preprocessing can already consume more power than the actual GPU trainers themselves in Meta's datacenters.}
This directly constrains training capacity due to fixed datacenter power budgets~\cite{barroso_datacenter}.
Finally, steady innovation in model complexity and DSAs for training are increasing data storage and bandwidth demands.
Figure~\ref{fig:storage_size} shows how industry-scale dataset sizes and online data ingestion bandwidth requirements have grown by over $2\times$ and $4\times$ over the past two years, respectively.
Barring similar innovation for DSI, we expect DSI infrastructures to severely limit training capacity at the datacenter-scale as training DSAs continue to yield higher performance per watt.

\input{fig-storage-size}
\input{tbl-takeaways}

In this paper, we present Meta's end-to-end DSI pipeline, which enables large-scale ML model training at scale.
Training data is generated by extract-transform-load (ETL) jobs that transform unstructured feature data and event logs collected across production fleets into structured training samples.
Petabyte-scale datasets are held in a centralized data warehouse as Hive tables~\cite{vldb09-hive}, which are subsequently stored as optimized columnar files in Tectonic~\cite{fast21-tectonic}, Meta's append-only distributed filesystem.
Finally, to handle intense online preprocessing demands, we present a production-deployed disaggregated online preprocessing framework called \textit{Data PreProcessing Service (\SystemName)} that iteratively reads and transforms mini-batches of training data, scaling from 10s to 100s of preprocessing nodes for each training job.

A key contribution of this paper lies in the deep system performance characterization for Meta's production-deployed DSI pipeline, supporting large-scale DNN training.
We describe Meta's collaborative feature engineering and model training process for DLRMs, which reveals key system design requirements and optimizations for Meta's underlying global DSI infrastructure.
Features for DLRM training evolve rapidly in our datasets, and samples are constantly generated.
This requires us to store and serve \textit{massive and dynamically-changing feature sets}, representing exabytes of cumulative storage.
Each training job requires an \textit{online preprocessing step, demanding significant compute, network, and memory resources}, in order to extract, transform, and load samples into materialized tensors for training.

Table~\ref{tbl:takeaways} summarizes the design principles of our DSI pipeline presented in Section~\ref{sec:dsi}, and key takeaways and open research problems for the wider community that we present in Section~\ref{sec:future}.
It connects each to specific system characterization results we present throughout Sections \ref{sec:training} to \ref{sec:preprocessing} for the production-deployed DSI pipeline.

In summary, our primary contributions are:

\begin{itemize}
\item We describe and identify the DSI pipeline as a critical and equally important, yet vastly understudied, component of datacenter-scale ML training infrastructures.
\item We provide an end-to-end description and share the design rationales behind our production-deployed DSI pipeline architecture, tailor-made to meet important requirements of DLRM training at scale.
\item We perform a deep characterization of industry-scale DLRM training workloads (summarized in Table~\ref{tbl:takeaways}), including coordinated training, data generation and storage, and online preprocessing, on our production hardware --- identifying critical bottlenecks and key insights.
\item We provide our outlook of important and open research questions for systems and computer architects to design and scale the DSI pipeline for large-scale training.
\end{itemize}

We hope that this work distills meaningful architecture and systems challenges in ML, beyond just DSAs for DNN training, and will guide the community to identify and focus on DL workloads that are representative of industry uses.

%% file: fig-ingestion-power.tex
\begin{figure}[t]
  \centering
  \includegraphics[width=3.33in]{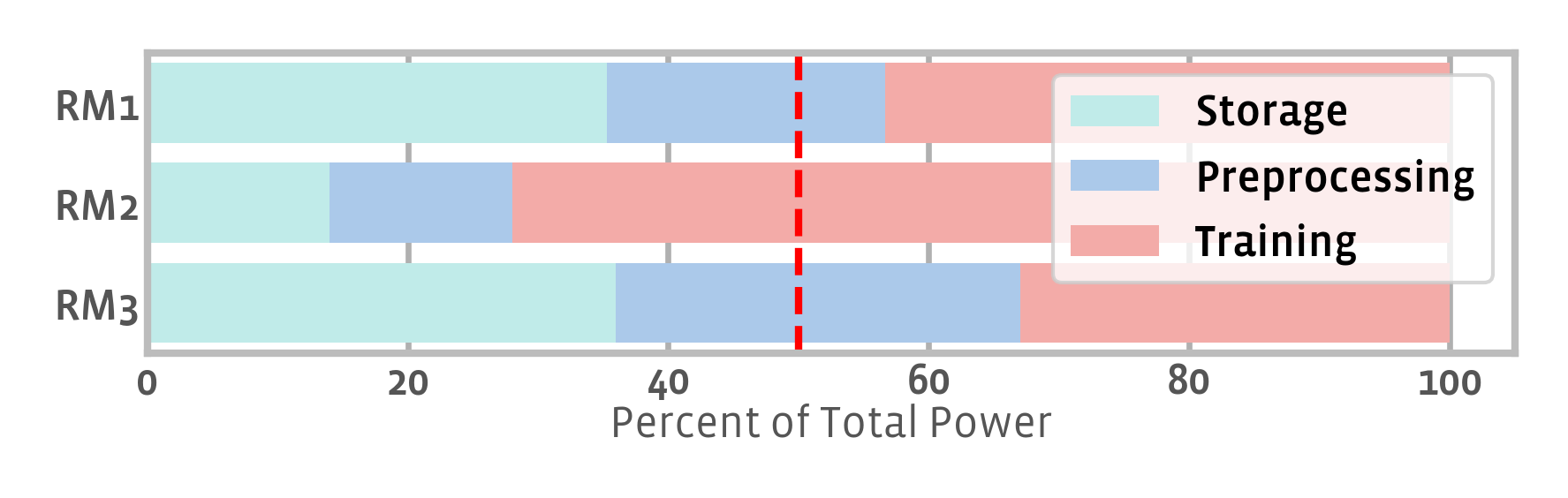} 
  \vspace{-10mm}
  \caption{\small Percent of storage, preprocessing, and training power required to train three production DLRMs, with line drawn at $50\%$. DLRMs exhibit diverse DSI resource requirements and can consume more power than training.}
  \vspace{-5mm}
  \label{fig:ingestion_power}
\end{figure}

%% file: fig-storage-size.tex
\begin{figure}[t]
  \centering
  \includegraphics[width=3.33in]{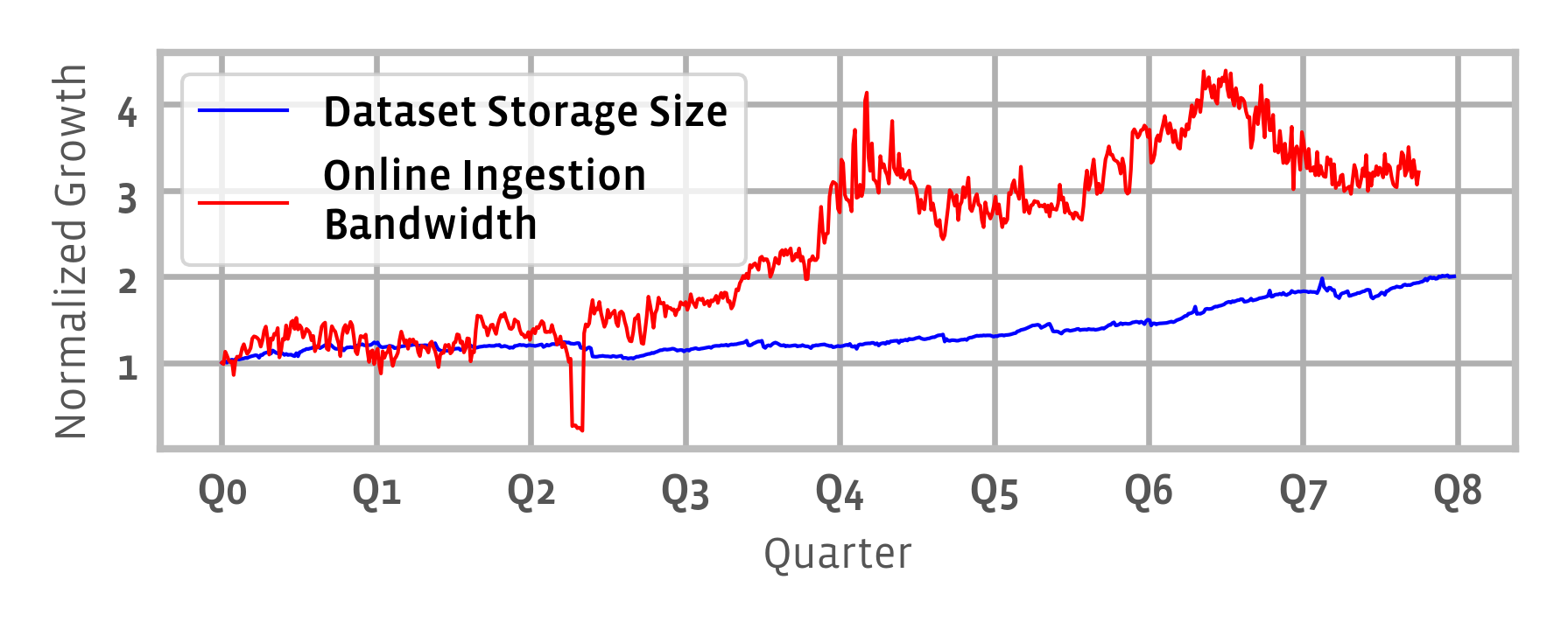} 
  \vspace{-10mm}
  \caption{\small Normalized training dataset size and online data ingestion bandwidth across our recommendation models. Storage and bandwidth is in the exabytes and hundreds of Tbps and has grown by over 2x and 4x over the past two years, respectively.}
  \label{fig:storage_size}
  \vspace{-5mm}
\end{figure}

%% file: tbl-takeaways.tex
\begin{table*}[t]
  \centering
  \caption{\small Summary of key takeaways and research opportunities, motivated by characterization results.}
    \label{tbl:takeaways}
  \vspace{-3mm}
  \footnotesize
  \resizebox{\linewidth}{!}{%
  \begin{tabular}{p{7cm}|p{9cm}}
  \hline
  
  \textbf{Key Lessons Learned and Takeaways} &  \textbf{Related Characterization Results ($\S$\ref{sec:training}-$\S$\ref{sec:preprocessing}}) \\ \hline
  \textbf{$\S$\ref{sec:dsi}:} Data storage and ingestion must be disaggregated to meet the large dataset capacity and online preprocessing requirements of training jobs.
  &
  \textbf{$\S$\ref{sec:training}:} Diverse training jobs run continuously in geo-distributed datacenters.
  \textbf{$\S$\ref{sec:dataset}:} Datasets exceed local storage capacity and are heavily filtered during reading.
  \textbf{$\S$\ref{sec:preprocessing}:} Using trainer CPUs for expensive online preprocessing results in data stalls.\\ \hline
  \textbf{$\S$\ref{sec:future-bottlenecks}:} There are significant compute, network, memory, and disk bottlenecks across storage, preprocessing, and trainer node hardware.
  These hardware bottlenecks are key areas of optimization for DSI.
  & 
  \textbf{$\S$\ref{sec:individual-dataset}:} Heavy feature filtering requires high IOPS from storage nodes.
  \textbf{$\S$\ref{sec:data-loading}:} Data loading at trainers requires considerable front-end trainer host resources. 
  \textbf{$\S$\ref{sec:extract_transform}:} Data extraction and transformation will be increasingly constrained by memory bandwidth.\\ \hline
  \textbf{$\S$\ref{sec:future-heterogeneous}:} Using and designing heterogeneous hardware for dataset storage systems and data ingestion can address many hardware bottlenecks, but require overcoming key challenges that arise in DSI pipelines.
  &
  \textbf{$\S$\ref{sec:reuse-dataset}:} Training jobs reuse popular features which can be cached in high IOPS/W storage.
  \textbf{$\S$\ref{sec:data-loading}:} Data loading requires expensive ``datacenter tax" operations.
  \textbf{$\S$\ref{sec:transformation-operations}:} We identified common transforms, especially feature generation, that are highly-resource intensive on CPUs. \\ \hline
  \textbf{$\S$\ref{sec:future-datacenter}:} 
  Designing datacenters for ML training require carefully provisioning compute, network, and storage capacity based on training jobs' diverse DSI requirements.
  Intelligent global training job schedulers can further improve efficiency across geo-replicated datacenters.
  &
  \textbf{$\S$\ref{sec:collaborative}:} Industry-scale training requires a collaborative training process across many models. \textbf{ $\S$\ref{sec:global-training}:} Each model requires periodic combo jobs with high concurrent compute demands that must be co-located with storage and scheduled across global regions. \\ \hline
  
  \textbf{$\S$\ref{sec:future-benchmarks}:} 
  Benchmarks are vital for guiding research directions.
  Current public datasets are not representative of industry; there is need for research in developing benchmark datasets.
  We identify important characteristics of industrial datasets.
  &
  \textbf{$\S$\ref{sec:feature-engineering}:} Industrial datasets are constantly updated with new data and features.
  \textbf{$\S$\ref{sec:individual-dataset}:} They are PB-sized and stored as structured samples in distributed file systems.
  \textbf{$\S$\ref{sec:individual-dataset}:} Training jobs read one epoch with feature-wise and row-wise filtering. \\ \hline
  \textbf{$\S$\ref{sec:future-optimization}:} DSI power constrains training capacity --- systems efficiency optimizations are vital. Efficiency optimizations must be co-designed across hardware/software and the end-to-end DSI pipeline. We walk through recent optimization examples.
  &
  Optimizations must consider application characteristics (selective reading - \textbf{$\S$\ref{sec:individual-dataset}}, feature popularity - \textbf{$\S$\ref{sec:reuse-dataset})}, hardware performance and bottlenecks (HDD IOPS - \textbf{$\S$\ref{sec:individual-dataset}}, preprocessing memory bandwidth - \textbf{$\S$\ref{sec:extract_transform}}), and trade-offs across end-to-end DSI pipeline systems (\textbf{$\S$\ref{sec:dsi}}). \\ \hline
  \end{tabular}
  } %
  \vspace{-3mm}
\end{table*}

%% file: background.tex
\section{Recommendation Model Background}\label{sec:background}
Personalized recommendation models are used to suggest new, relevant content to users to provide meaningful interactions.
These models are highly potent across a breadth of tasks.
They leverage features from a user and a potential recommendation, and output a prediction (e.g., click-through rate) of how likely the user is to interact with the recommendation.
For example, a video hosting site may use a user's set of liked videos and candidate videos' genres to rank new videos to recommend to the user.

Recommendation models are trained using mini-batch stochastic gradient descent (SGD)~\cite{gradient-descent}, similar to most vision and natural language processing (NLP) models.
SGD generalizes a model to complex distributions by iteratively updating the model's weights to minimize a loss function, given successive \textit{mini-batches} of samples.
Each training sample is represented as a preprocessed tensor of \textit{features} and a label.

\input{fig-pipeline-bd}

Our production recommendation models are built on the open-source DLRM architecture~\cite{dlrm}.
Modern DLRMs are massive, consisting of over 12 trillion parameters to train, requiring $\approx1$ zetaFLOPs of total compute~\cite{arxiv-mudigere}.
To meet the compute requirements of DLRM training, we built the ZionEX hardware platform.
Each node contains 8 NVIDIA A100 GPUs connected with NVLink for intra-node communication.
Each GPU has a dedicated 200 Gbps RDMA over Converged Ethernet (RoCE) NIC for inter-node communication over a dedicated backend network, connecting thousands of GPUs to form a datacenter-scale AI training cluster.
A node also contains four CPU sockets, each with a dedicated 100 Gbps NIC connected to our regular datacenter network for data ingestion.
We defer readers to \cite{arxiv-mudigere} for more details on ZionEX.

To fully leverage our training hardware, we use data~\cite{arxiv-krizhevsky} and model~\cite{nips12-dean} parallelism by replicating and sharding the model across multiple ZionEX training nodes.
We distribute different mini-batches to each trainer.
Trainer nodes synchronize embeddings, activations, and gradients with each other using collective communication primitives over the backend network, iterating until a certain model quality metric (e.g., normalized entropy~\cite{adkdd14-he}) is reached.
At Meta, we use hundreds of distinct recommendation models in production across our services.
We continuously train new versions of each model offline (Section~\ref{sec:training}), and  update production models periodically~\cite{hpca18-hazelwood}.

Each training job relies on a data storage and ingestion (DSI) pipeline to supply each trainer with training data throughout the duration of the job.
The DSI pipeline is thus responsible for \textit{generating} training samples, \textit{storing} them into datasets, and \textit{preprocessing} samples into tensors loaded in device memory, i.e., GPU HBMs.

%% file: fig-pipeline-bd.tex
\begin{figure*}[t]
  \includegraphics[width=\textwidth]{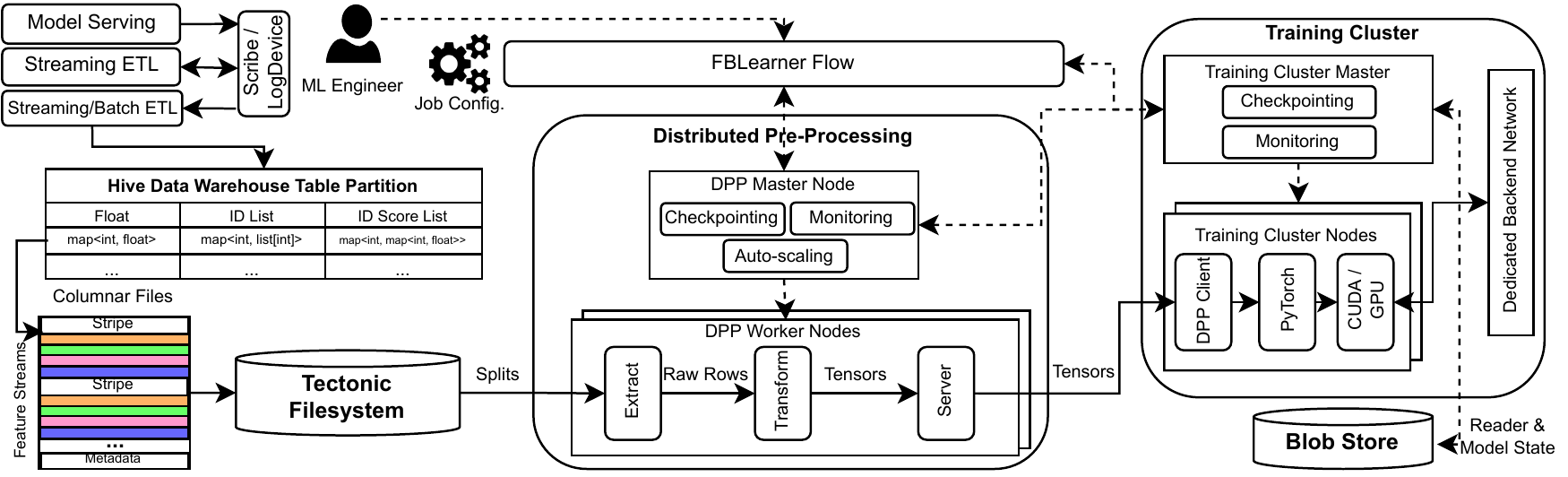}
  \vspace{-5mm}
  \caption{\small Production data storage and ingestion pipeline architecture. Solid and dashed lines represent data and control flow, respectively.}
  \label{fig:pipeline_bd}
  \vspace{-3mm}
\end{figure*}

%% file: dsi.tex
\section{Disaggregated Data Storage, Ingestion, and Training Pipeline}\label{sec:dsi}
In this section, we present our DSI requirements and the storage, preprocessing, and training systems that compose Meta's DSI pipeline as shown in Figure~\ref{fig:pipeline_bd}.

\subsection{Data Generation and Storage}\label{sec:distributed-storage}
\noindent\textbf{Overview and Requirements.}
Figure~\ref{fig:pipeline_bd} shows how fresh training samples are continuously generated by our model serving framework in order to ensure model accuracy and comply with privacy requirements~\cite{hpca18-hazelwood}.
Each sample is created as services evaluate a user and item using the model serving framework.
The framework first generates an extensive set of \textit{features} (e.g., a user's liked pages) as input to an appropriate model, which outputs a prediction used for recommendation tasks.
The requesting service then monitors \textit{events} representing the outcome of each recommendation (e.g., if a user interacted with a post).
These features and events are logged at serving time to avoid data leakage~\cite{kdd11-kaufman} between model serving and training.
Subsequent streaming and batch extract-load-transform (ETL) jobs continuously join and label raw feature and event logs into labeled and schematized samples.

Training samples are placed in a storage solution that must meet several key requirements.
First, individual tables require \textit{tens of thousands of features, with features being constantly added or removed.}
Training jobs must be able to \textit{dynamically and selectively read} stored features.
Second, \textit{developer productivity is paramount}.
We must allow ranking engineers and the underlying infrastructure to easily work across hundreds of models and tables via a centralized data warehouse using a common schema.
Furthermore, ranking engineers frequently run interactive queries using Spark~\cite{nsdi12-spark} or Presto~\cite{icde19-presto} as a part of feature engineering in addition to training.
Finally, \textit{datasets are continuously updated with fresh samples}.

\subsubsection{Data Generation}~\label{sec:generation}
To extract and aggregate billions of features and events across the entire fleet each day, we rely on Scribe~\cite{scribe} --- Meta's global distributed messaging system.
Each service responsible for serving models or handling interactions continuously passes raw feature and event logs to a Scribe daemon running on every host.
Scribe then groups logs into record-oriented logical streams and stores each stream into LogDevice~\cite{logdevice} --- a reliable distributed store for append-only, trimmable streams built on top of RocksDB~\cite{rocksdb}.

To update production models, streaming engines first join and label raw feature and event logs from Scribe and publish labeled samples into various Scribe streams used to update in-production models. %
Various streaming and batch processing engines, such as Spark~\cite{nsdi12-spark}, further join, label, and filter samples from Scribe streams to produce partitioned (e.g., hourly or daily) offline datasets used to train new production model versions.
Because traditional engines work well to generate training data, require comparatively little compute and power resources to dataset storage and ingestion, and are not on the critical path of training, we elide in-depth discussion here.

\subsubsection{Data Storage}~\label{sec:warehouse}
We store training samples in a data warehouse as partitioned Hive~\cite{vldb09-hive} tables because of Hive's compatibility with both internal systems and open source engines including Spark and Presto.
Samples are represented as structured rows, each containing features and labels, with features requiring the vast majority ($>99\%$) of stored bytes.
To ensure interoperability of the DSI pipeline across hundreds of models and tens of thousands of features, we store two types of features, dense and sparse, in map columns.
A dense feature column maps a feature ID to a continuous value (e.g., current time).
A sparse feature column maps a feature ID to a \textit{variable length} list of categorical values (e.g., page IDs).
Some sparse features are stored in an additional column that further associates each categorical value with a floating point "score" used for weighing (e.g., page creation time).

We encode Hive tables in a columnar file format (DWRF), forked from Apache ORC~\cite{apache-orc}.
Like ORC, rows are stored across multiple files.
Each file contains a set of stripes, representing a number of table rows.
Stripes are further divided into compressed and encrypted streams.
A key distinction of DWRF is the ability to enable feature filtering at the storage layer by flattening each feature column and storing features as thousands of separate logical columns at the file layer (see Section~\ref{sec:future}).
Each flattened feature column is subsequently encoded as one or more streams, depending on its schema and encoding.
Stripes are periodically flushed and appended to the file.
Files are written in Tectonic~\cite{fast21-tectonic} --- Meta's exabyte-scale distributed append-only filesystem.
Tectonic splits files into durable blocks distributed across HDD storage nodes. %

\subsection{Online Preprocessing}
\noindent\textbf{Overview and Requirements.}
Each training job uses an online (training-time) preprocessing pipeline to continuously transform raw samples in storage into \textit{preprocessed} tensors in a trainer's memory. %
Like offline data generation, online preprocessing is commonly subdivided into ETL phases.

Raw bytes are \textit{extracted} from storage and decoded into training samples, a process involving filtering, decryption, decompression, reconstruction, and other format transformations.
Training samples are next \textit{transformed} into tensors.
Float values may be normalized, and categorical values may be hashed, clipped, or even sorted.
New features may even be derived from multiple raw features. %
Once features are preprocessed, they are batched together into tensors.
The tensors are \textit{loaded} into trainers, usually in device memory (e.g., HBM of GPUs).

Online preprocessing has distinct requirements that differ from those met by traditional ETL engines. %
First, transformations for online preprocessing are localized to each mini-batch, not across many rows. %
Second, online preprocessing is on the critical path of training and must match throughput required by trainers.
Right-sizing online preprocessing throughput is critical to avoid either over-provisioning resources or introducing data stalls~\cite{vldb21-mohan} that will bottleneck and under-utilize expensive trainers. 
Finally, online preprocessing runs concurrently alongside \textit{each} training job, requiring significant cumulative compute and power resources that scales with training capacity, unlike offline data generation.

\subsubsection{Scalable Preprocessing with \SystemName}\label{sec:dpp}
\textit{Data PreProcessing Service} (\SystemName) is our disaggregated service that provides online preprocessing for training jobs across the datacenter fleet.
\SystemName is responsible for reading raw training data from storage, preprocessing it into ready-to-load tensors, and supplying the tensors to each training node's PyTorch~\cite{neurips19-pytorch} runtime.
We designed \SystemName to both \textit{scale to right-sized resources and eliminate data stalls across disparate jobs}, as well as \textit{enable vital productivity mechanisms for developers and ML engineers}.
We meet these requirements by dividing \SystemName into a data and control plane, which are designed to enable application throughput and ease-of-use, respectively.
The control plane consists of a \SystemName Master, and the data plane consists of \SystemName Workers and Clients.
As shown in Figure~\ref{fig:pipeline_bd}, ML engineers launch training jobs via FBLearner Flow~\cite{fblearner-flow}, which then launches a \SystemName Master and at least one \SystemName Worker on general-purpose compute nodes.

\textit{\textbf{\SystemName Control Plane.}}
At the beginning of each training job, the \SystemName Master receives a session specification (a PyTorch \textsc{DataSet}) that reflects the preprocessing workload, containing the dataset table, specific partitions, required features, and transformation operations for each feature.
The \SystemName Master enables scalable work distribution by breaking down the entire preprocessing workload, across petabytes of data, into independent and self-contained work items for the data plane called \textit{splits} that represent successive rows of the entire dataset.
The Master serves splits to \SystemName Workers upon request and tracks progress as splits are completed.

In addition to work distribution, the \SystemName Master is responsible for fault tolerance and auto-scaling.
The \SystemName Master periodically creates a checkpoint which can be used to restore reader state on failure.
The \SystemName Master also continuously monitors Worker health, automatically restarting any Workers that have failed without needing a checkpoint restore due to Workers' stateless design. %
The \SystemName Master itself is replicated to avoid being a single point of failure.

Finally, the \SystemName Master implements auto-scaling via a controller.
The controller collects utilization (CPU, memory, and network) statistics and the number of buffered tensors from each \SystemName Worker.
It then periodically evaluates scaling decisions, calculating the number of \SystemName Workers to either drain or launch with the goal of maintaining a non-zero number of buffered tensors (indicating that trainer demand is met) and maximum CPU, network, and memory utilization.
In doing so, the \SystemName Master eliminates data stalls with minimal \SystemName resource requirements.

\textit{\textbf{\SystemName Data Plane.}}
\SystemName Workers and Clients are responsible for data plane operations of \SystemName.
Workers are designed to effortlessly scale out to eliminate data stalls.
Workers are stateless, precluding any limit to how many Workers can exist in a given \SystemName session.
They only communicate with the \SystemName Master (to fetch work items) and a limited number of \SystemName Clients (to serve tensors); all transformations within a mini-batch are performed locally.
On startup, each Worker pulls a set of transformations from the Master, represented by a serialized and compiled PyTorch module, that it will use during preprocessing.
Workers then continuously query for and process splits from the \SystemName Master.

As shown in Figure~\ref{fig:pipeline_bd}, each split requires Workers to extract, transform, and (partially) load training data.
Specifically, Workers begin by reading, decompressing, and decrypting raw Tectonic chunks.
Sets of raw chunks are then reconstructed into streams and decoded into raw table rows, filtering out unused features if necessary.
Next, it applies the specified transformations to each raw feature using high-performance C++ binaries. %
Once features are transformed, Workers batch samples together into tensors to be loaded onto GPU trainers.
We ensure that transient delays in the pipeline do not introduce data stalls by maintaining a small buffer of tensors in each Worker's memory.

\textbf{\textit{Trainers.}}
\SystemName Clients form the other half of the data plane.
A Client runs on each training node, exposing a hook that the PyTorch runtime can call to obtain preprocessed tensors.
These requests are transparently transformed into a simple RPC request which returns a batch of tensors from the Worker buffer.
By offloading computationally expensive operations to \SystemName Workers and enabling Client multithreading, Clients do not bottleneck the data ingestion pipeline.
To ensure that Client and Worker network connections can scale, each Client uses partitioned round robin routing, capping the number of connections that Clients and Workers need to maintain.

To enable large-scale recommendation model training, we have built high-performance training clusters~\cite{arxiv-mudigere}, enabling individual jobs to train on hundreds to thousands of GPUs.
Each training job is controlled by a Trainer Master, which manages the overall training session.
On each trainer, a PyTorch runtime manages the local training workflow, transferring preprocessed tensors between the \SystemName Client and GPU device memory.
Parameter updates between trainers occur over a dedicated backend network and do not impact data ingestion.

%% file: training.tex
\section{Coordinated Training at Scale}\label{sec:training}
Next, we explore how DLRMs are trained and deployed at Meta.
We do so because large-scale training deployments require thousands of training jobs over hundreds of models and datasets, all running on a shared global infrastructure.
Understanding industrial training jobs highlights important system and infrastructure implications not found in  commonly-studied hyperparameter (HP) tuning or isolated training jobs. %
We focus on the three representative recommendation models (RMs) highlighted in Figure~\ref{fig:ingestion_power}, denoted $RM_{1,2,3}$, as these models are the most widely-used and training resource-intensive.

\subsection{Collaborative Release Process}\label{sec:collaborative}
Hundreds of recommendation models are deployed in production at Meta, each continuously developed by many training jobs.
Each training job produces a new model version with the goal of becoming the next production model.
Many of our models are supported by large teams of ranking engineers, requiring the need to allow continuous experimentation across engineers while avoiding conflicts between model versions and conserving limited training capacity.
This need naturally arises as model engineering teams mature~\cite{mastercook}.

We thus adopted a regimented release process which occurs over three phases.
First, ML engineers \textit{explore} their ideas (e.g., new features or model architectural improvements) on top of the current production model through hundreds to thousands of small training jobs.
Exploratory jobs generally require less compute and use a small fraction (typically $<5\%$) of its respective table's total samples.
Next, the most promising ideas are periodically \textit{combined} in various permutations to generate tens to hundreds of training jobs.
These \textit{combo jobs} are large and trained within a short window, demanding immense parallelism and the majority of the table.
Finally, the most promising \textit{release candidates} (RCs) are further trained and evaluated on fresh data, and the most accurate model is deployed in production.
While these jobs are large, there are only a few.
\input{fig-combo-skew}

Counter-intuitively, the model release process can result in \textit{more} diversity in terms of temporal locality, model architectures, and feature sets than in isolated or HP tuning jobs~\cite{fast20-quiver, hotcloud19-oneaccess}. 
This is because training capacity is highly-constrained compared to per-job compute requirements, requiring engineers to combine many architecture and feature proposals into one combo job.
Figure~\ref{fig:combo_skew} illustrates how the model design space is explored given compute constraints.
While individual jobs are long-running and can take over 10 days to train, many jobs fail or are killed because their performance is lackluster.
Instead of waiting to launch jobs synchronously, engineers will immediately schedule new jobs to maximize the number of explored ideas within the combo time window, resulting in a large temporal skew between jobs.

\subsection{Global Training Utilization}\label{sec:global-training}
\input{fig-compute}
\input{fig-model-compute}

The aforementioned collaborative training jobs, including exploratory, combo, and release candidates, run on the global fleet of training (including DSI) infrastructure, spread across global regions, each with multiple datacenters~\cite{hpca18-hazelwood}.
Figure~\ref{fig:compute} shows a historical utilization, in terms of normalized compute, of all collaborative training jobs across DLRMs over one calendar year across our entire fleet.
We observe distinct peaks in utilization, corresponding to periods where many models concurrently train combo jobs.
Because these combo jobs are on the critical path of model release, we must explicitly architect our datacenters with sufficient storage, preprocessing, and training capacity to meet the peak utilization of combo jobs.

Furthermore, as we explore in Section~\ref{sec:dataset}, each model reads a distinct dataset.
At the same time, cross-region (and often cross-datacenter) bandwidth is highly-constrained.
This requires systems and datacenter architects to co-locate DSI resources with trainers themselves and provision enough capacity for each. %
Figure~\ref{fig:model_compute} shows a bar chart of the relative compute demand of the ten most commonly-run models, broken down by the region in which they ran.
Our global scheduler currently balances training jobs for each model across regions, requiring each region to contain a copy of all models' datasets.
Bin-packing opportunities can reduce storage costs, with care to ensure data availability for each model as its peak compute demand can exceed regional capacity.

\subsection{Feature Engineering}\label{sec:feature-engineering}
\input{tbl-feature-lifecycle}
The sets of features stored to a dataset and read by training jobs can also vary heavily, as features also undergo rapid experimentation and productionization. %
To understand feature storage variability, Table~\ref{tbl:feature_lifecycle} shows the total number of new features proposed for $RM_1$'s production dataset within a 6-month window and the status of the feature 6 months later.
Beta features are not actively logged, but may be back-filled or injected (i.e., dynamically joined) for each exploratory training job.
Experimental features are used as a part of combo or release candidate jobs.
If the release candidate job becomes the next production model, its used features become active, while some older features may become deprecated following a review process or even reaped to protect user privacy.
Experimental, active, and deprecated features are actively written to the dataset. %
We observe that features are rapidly changing in production datasets, with hundreds of new features added and deprecated each month.
Thus, efficient ML data storage infrastructure must adapt to frequent changes in the feature set.

\subsection{Summary of Key Takeaways}
Training production models requires a collaborative release process across hundreds of engineers.
Critically, ideas are periodically amalgamated in a large number of concurrent combo jobs for each model, resulting in large peaks in training and DSI resources across our fleet during this phase.
Because combo jobs are on the critical path of model release, we must design datacenters with sufficient capacity across global regions for peak demand corresponding to combo jobs. %
This capacity required is spread across hundreds of models with varying compute demand, motivating the need for efficient co-location and scheduling of jobs and datasets across regions to reduce inter-region storage and network demands.
Finally, we explored how training jobs are temporally skewed and exhibit diverse model architectures, and datasets are continuously evolving, inhibiting system optimizations that assume highly-synchronized and similar training jobs or static datasets, e.g. \cite{fast20-quiver, vldb21-mohan, hotcloud19-oneaccess}.

%% file: fig-combo-skew.tex
\begin{figure}[t]
  \centering
  \includegraphics[width=3.33in]{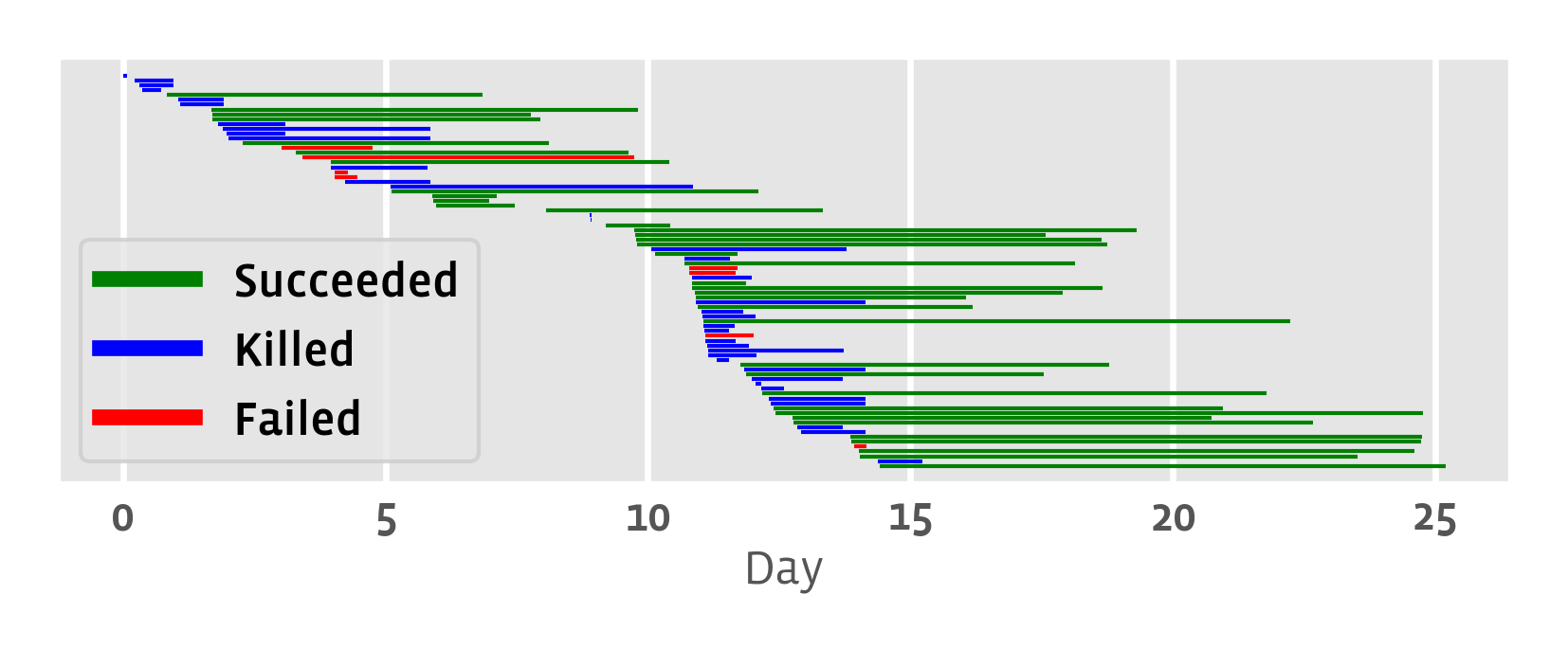} 
  \vspace{-8mm}
  \caption{\small Chart showing skewed and variable training duration and status of 82 $RM_1$ combo jobs within one model release iteration.}
  \vspace{-4mm}
  \label{fig:combo_skew}
\end{figure}

%% file: fig-compute.tex
\begin{figure}[t]
  \centering
  \includegraphics[width=3.33in]{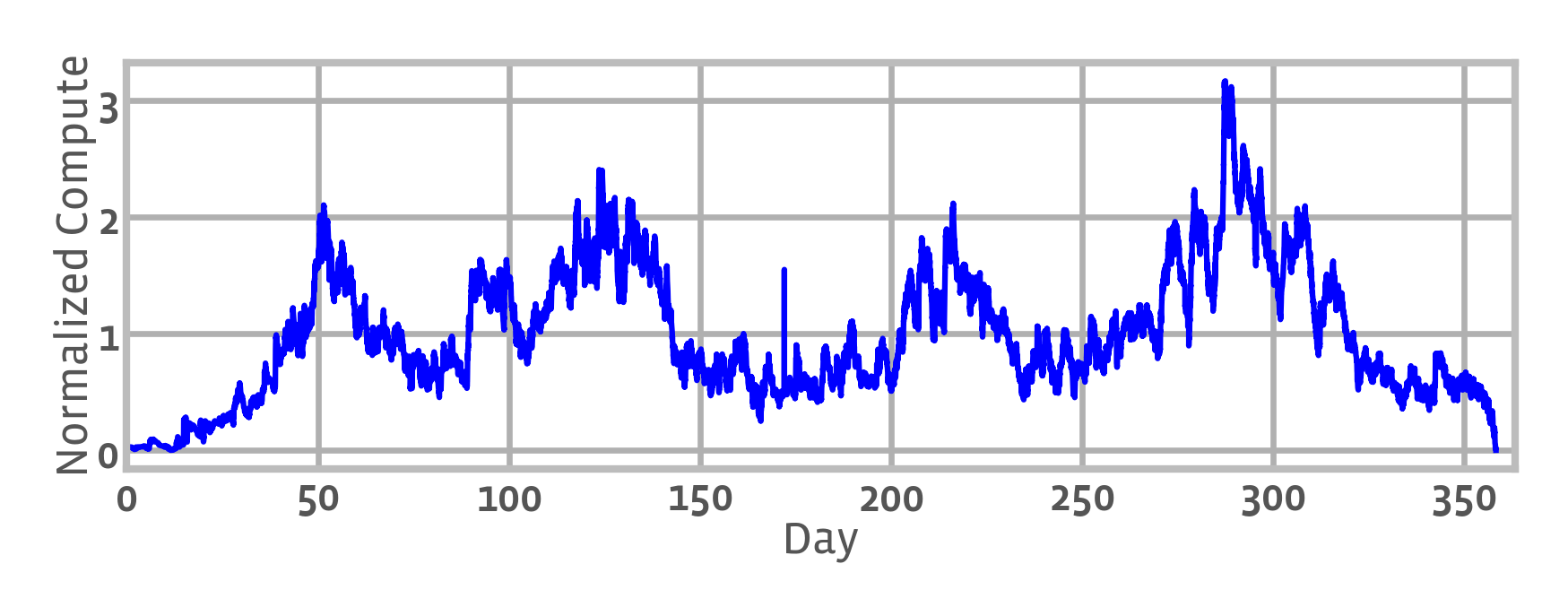} 
  \vspace{-8mm}
  \caption{\small Normalized daily peak compute utilization over all collaborative training jobs over one year, showing peaks corresponding to combo jobs.}
  \label{fig:compute}
  \vspace{-4mm}
\end{figure}

%% file: fig-model-compute.tex
\begin{figure}[t]
  \centering
  \includegraphics[width=3.33in]{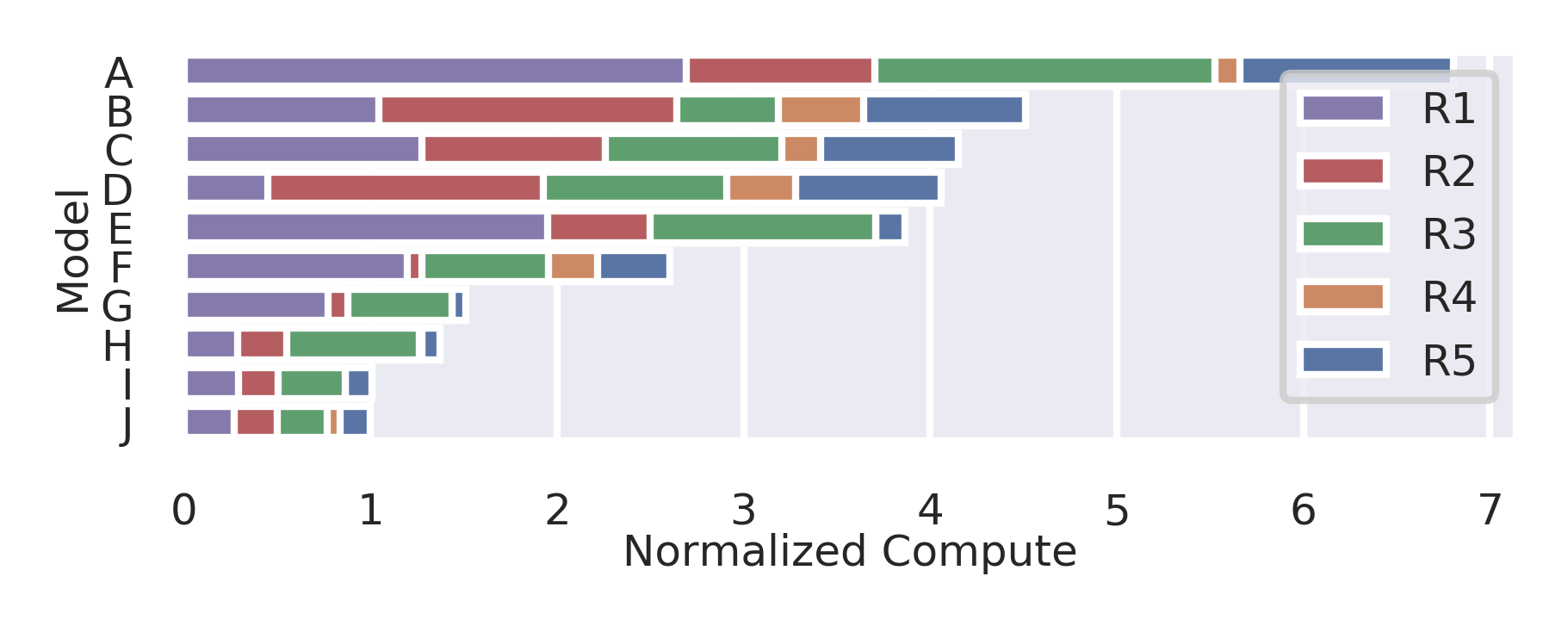}
  \vspace{-8mm}
  \caption{\small Compute demand over ten most commonly-used models (A-J), split by global region (R1-R5), normalized to model J.}
  \label{fig:model_compute}
  \vspace{-3mm}
\end{figure}

%% file: tbl-feature-lifecycle.tex
\begin{table}[t]
\centering
\footnotesize
\caption{\small Number of features created for $RM_1$ dataset within a 6 month window and their status 6 months later.}
\label{tbl:feature_lifecycle}
\vspace{-3mm}
\begin{tabular}{@{}rrrrr@{}}
\toprule
Beta  & Experimental & Active & Deprecated & Total \\ \midrule
10148 & 883          & 1650   & 1933       & 14614 \\ \bottomrule
\end{tabular}%
\vspace{-3mm}
\end{table} %

%% file: dataset.tex
\section{Understanding Data Storage and Reading}\label{sec:dataset}
We next explore how datasets are stored in our data warehouse and read by training jobs, highlighting implications to our storage hardware and infrastructure.

\subsection{Individual Jobs Read and Filter Large Datasets}\label{sec:individual-dataset}
Benchmark datasets are typically re-read multiple times (epochs) to reach target accuracy~\cite{mlsys20-mlperf}.
Thus, existing work focuses on randomly modifying~\cite{neurips20-randaugment, cvpr19-autoaugment, arxiv-choi, neurips20-agarwal, atc21-revamper} or caching~\cite{tfdata, vldb21-mohan, fast20-quiver, arxiv-yang} data across epochs to improve DSI efficiency.

Unlike these benchmarks, production training jobs are not constrained by the amount of data.
Instead, model size and compute capacity limits constrain the number of features and samples each job can use, respectively. %
Production training jobs do not require stochastic preprocessing across multiple epochs, but instead can reach a desired target accuracy with (less than) one epoch containing many samples.
Individual training jobs specify a given table as its dataset, along with filters that select a subset of data within the table along two dimensions: a variable number of partitions (row filter) and a set of features within each sample to read (column filter).

\input{tbl-table-size}

We begin by analyzing one representative RC training job from $RM_{1,2,3}$.
Table~\ref{tbl:table_size} shows the (compressed) size characteristics of each model's respective production training data table.
It also shows the size of each partition in the table (partitioned by date) and the cumulative size of the partitions used by the training job for each $RM$.
Even our largest training jobs often read less than the entire available dataset and each sample only once.
The partitions that are read still require petabytes of data, which is significantly larger than the local storage capacity at each trainer node, contrary to prior assumptions~\cite{vldb21-mohan}.
Furthermore, as shown in Figure~\ref{fig:storage_size}, dataset sizes for production models are continuously growing, driven by multiple factors such as organic user growth, reduced downsampling, and an increase in engineered features.

\input{tbl-model-characteristics}
\input{tbl-dataset-characteristics}

Next, we study how a training job selects data along the feature (column) dimension.
Individual training jobs specify a \textit{feature projection}, consisting of a list of desired features to be read from all rows in the designated partitions.
Table~\ref{tbl:model_characteristics} shows the number of dense, sparse, and derived features required by a representative RC model version for each $RM$.
These model versions require $504-1221$ and $42-306$ dense and sparse features, respectively.
This is in contrast to Table~\ref{tbl:dataset_characteristics}, which shows that significantly more features are logged in each model's table.
Each training job only needs to read $9-11\%$ of stored features.
Even when accounting for the number of bytes read, Table~\ref{tbl:dataset_characteristics} highlights that $RM_{1,2,3}$ only read between $21$ and $37$ percent of stored bytes across used partitions.
The relative increase in read bytes is because read features typically exhibit larger coverage (i.e., fraction of samples logging the feature) and sparse feature lengths, and thus require more bytes, as these features contribute stronger signals to model quality and are thus favored by ML engineers.

While there appears to be room to reduce feature collection, feature experimentation is essential for ML engineers.
We prioritize developer productivity and heavily err on the side of keeping features, even at the cost of storage, to ensure that ML engineers have access to the features they need. %

\input{tbl-io-size}
Selective reading also has further implications for the performance of our storage nodes.
Table~\ref{tbl:io-size} shows the distribution, in bytes, of a representative $RM_1$ training job's I/O sizes from storage.
Heavy filtering and columnar storage of features on disk (Section~\ref{sec:dsi}) results in relatively-small contiguous regions for read features.
Further software-hardware co-design is needed to ensure that disk seeks do not cripple storage IOPS.

\subsection{Data is Reused Across Training Jobs}\label{sec:reuse-dataset}
\input{fig-reuse-analysis}
While individual training jobs require extensive filtering, training jobs do collectively reuse data.
Inter-job data reuse can occur throughout the model release process because ML engineers do not develop an entirely new model architecture and feature set each iteration, but instead largely build upon a common baseline (e.g., the current production model version).

Figure~\ref{fig:reuse_analysis} shows that, based on training runs for $RM_{1,2,3}$ over one month, training runs for each model tend to favor specific bytes.
The x-axis shows a CDF of bytes within the model's used set of table partitions.
The y-axis shows the percent of all I/O from storage that the most-popular $x$ percent of stored bytes contribute to.
To serve $80\%$ of traffic from storage, we only require the most commonly-used $39$, $37$, and $18$ percent of $RM_1$'s, $RM_2$'s, and $RM_3$'s datasets, respectively.
Combined with Table~\ref{tbl:dataset_characteristics}, Figure~\ref{fig:reuse_analysis} also highlights how used features and bytes vary across training jobs.
RM3 exhibits little variance in features --- Table~\ref{tbl:dataset_characteristics} and Figure~\ref{fig:reuse_analysis} show that both individual and collective models read roughly $21\%$ of the stored bytes.
Meanwhile, $RM_1$ and $RM_2$ show high variance; individual jobs only read $37\%$ and $34\%$ of the stored bytes, respectively, while jobs collectively read over $60\%$ of the stored bytes.

\subsection{Summary of Key Takeaways}
In this section, we explored how features are stored in petabyte-scale datasets that greatly exceed local storage capacities, requiring training jobs to read samples from a centralized data warehouse (Section~\ref{sec:distributed-storage}).
Furthermore, each training job requires extensive filtering both in the number of samples (rows) and features extracted from each sample (columns).
Column-wise filtering results in small reads from storage because features are stored in columnar files.
Finally, across training jobs for a given model, we observed significant reuse in commonly-used features, with $40\%$ of bytes contributing to over $80\%$ of read throughput.

%% file: tbl-table-size.tex
\begin{table}[t]
  \centering
  \footnotesize
    \caption{\small Compressed sizes of all table partitions, each partition, and the cumulative partitions used by a representative release candidate training job for each RM.}
  \label{tbl:table_size}
  \vspace{-3mm}
  \begin{tabular}{@{}lrrr@{}}
    \toprule
    Model & All Partitions (PB) & Each Partition (PB) & Used Partitions (PB) \\ \midrule
    RM1   & 13.45           & 0.15                & 11.95             \\
    RM2   & 29.18           & 0.32                & 25.94             \\
    RM3   & 2.93            & 0.07                & 1.95              \\ \bottomrule
  \end{tabular}%
  \vspace{-4mm}
\end{table}

%% file: tbl-model-characteristics.tex
\begin{table}[t]
  \centering
  \footnotesize
  \caption{\small Feature characteristics of production models.}
  \label{tbl:model_characteristics}
  \vspace{-3mm}
  \begin{tabular}{@{}lrrr@{}}
    \toprule
    Model Class               & \# Dense Features & \# Sparse Features & \# Derived Features \\ \midrule
    \multicolumn{1}{l|}{RM1} & 1221              & 298                & 304                 \\
    \multicolumn{1}{l|}{RM2} & 1113              & 306                & 317                 \\
    \multicolumn{1}{l|}{RM3} & 504               & 42                 & 1                   \\ \bottomrule
  \end{tabular}
  \vspace{-3mm}
\end{table}

%% file: tbl-dataset-characteristics.tex
\begin{table}[t]
  \centering
  \footnotesize
  \caption{\small Dataset characteristics for each model.} %
  \label{tbl:dataset_characteristics}
  \vspace{-3mm}
  \begin{tabular}{@{}lrrrrrr@{}}
    \toprule
    Dataset & \thead{\# Float \\ Feats.} & \thead{\# Sparse \\ Feats.} & \thead{Avg. \\ Coverage} & \thead{Avg. \\ Sparse Feat. \\ Length} & \thead{\% Feats. \\ Used} & \thead{\% Bytes \\ Used} \\ \midrule
    \multicolumn{1}{l|}{RM1} & 12115 & 1763 & 0.45 & 25.97 & 11 & 37 \\
    \multicolumn{1}{l|}{RM2} & 12596 & 1817 & 0.41 & 25.57 & 10 & 34 \\
    \multicolumn{1}{l|}{RM3} & 5707  & 188  & 0.29 & 19.64 & 9  & 21 \\ \bottomrule
  \end{tabular}
  \vspace{-4mm}
\end{table}

%% file: tbl-io-size.tex
\begin{table}[t]
  \centering
  \footnotesize
\caption{\small I/O Sizes for features read by an RM1 training job.}\label{tbl:io-size}
\vspace{-3mm}
\begin{tabular}{@{}llllllll@{}}
\toprule
            & Mean   & Std   & p5 & p25 & p50    & p75    & p95    \\ \midrule
I/O Size (B) & 23.2K & 117K & 18 & 451 & 1.24K & 3.92K & 97.7K \\ \bottomrule
\end{tabular}%
\vspace{-3mm}
\end{table}

%% file: fig-reuse-analysis.tex
\begin{figure}[t]
  \centering
  \includegraphics[width=3.33in]{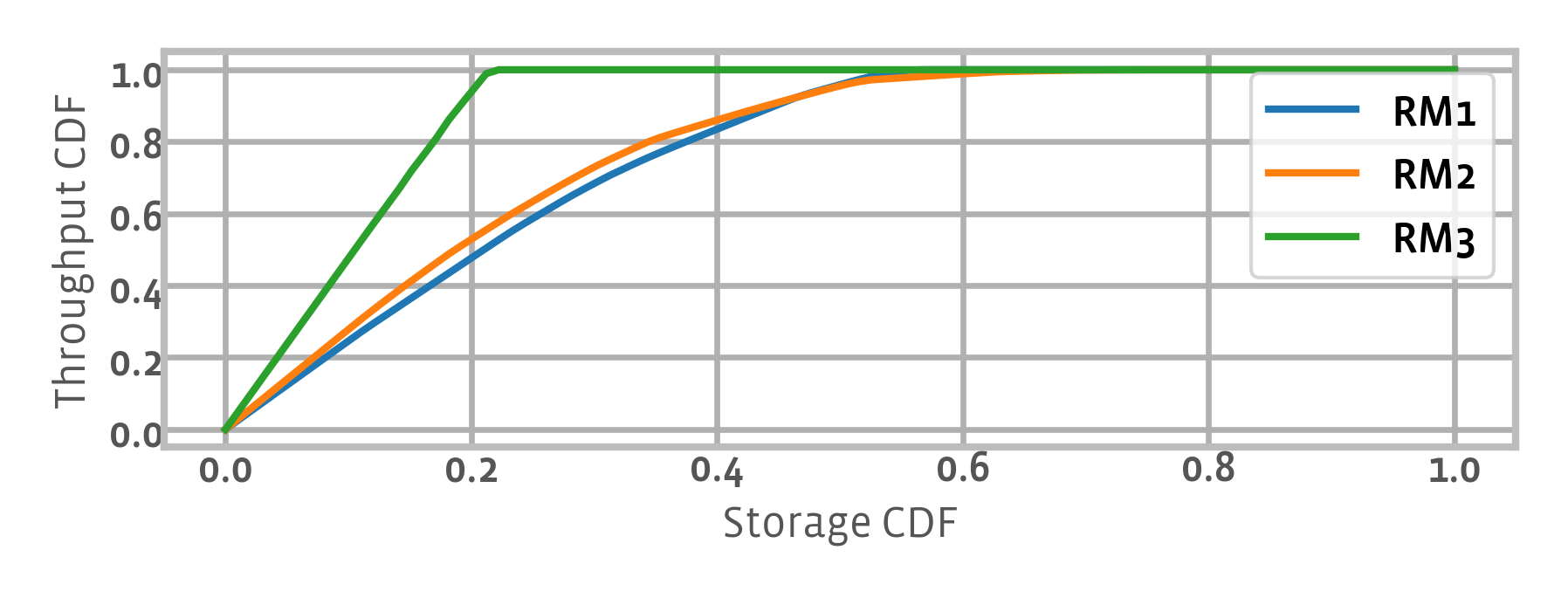}
  \vspace{-8mm}
  \caption{\small CDF of popular bytes to throughput absorbed, across one month of each RM's runs. Popular bytes are reused across runs.}
  \label{fig:reuse_analysis}
  \vspace{-4mm}
\end{figure}

%% file: preprocessing.tex
\section{Understanding Online Preprocessing}\label{sec:preprocessing}
\input{tbl-onbox}
Distributed trainers drive strict data ingestion bandwidth requirements.
Data stalls result when online preprocessing throughput is less than the aggregate throughput of the trainers themselves, underutilizing GPU resources~\cite{vldb21-mohan}.

Current preprocessing solutions, which perform preprocessing on the CPUs of each training node, can cause data stalls.
To demonstrate this, we ran a training job for $RM_1$ on a training node consisting of two 28-core x86 CPU sockets, two 100 Gbps frontend NICs for data ingestion, and a total of 8 NVIDIA V100 GPUs.
The trainer read from distributed storage, preprocessed each mini-batch using the production PyTorch~\cite{neurips19-pytorch} stack, and performed training on the same machine.
Table~\ref{tbl:onbox} shows that $56\%$ of GPU cycles were spent stalled waiting for training data.
The high CPU utilization shows that the trainer's CPUs cannot preprocess data fast enough to serve the GPUs, motivating us to build \SystemName to eliminate data stalls (Section~\ref{sec:dpp}).

We next seek to understand the preprocessing bottlenecks of our production DSI pipeline in detail.
To do so, we analyzed the preprocessing throughput demanded by GPUs across RMs.
We then traced through data loading resource requirements at GPU trainers, and data extraction and transformation resource requirements at \SystemName Workers.

\subsection{GPU Training Throughput}\label{sec:gpu-trainers}
\input{tbl-gpu-throughput}
We measured the online preprocessing throughput required by each $RM$ by running a production training job for each $RM$ on a training node, ensuring that GPUs were not stalled by serving tensors from an in-memory buffer.
Table~\ref{tbl:gpu_throughput} shows the per-trainer node GPU throughput requirements (i.e., tensor ingestion rate) for each representative $RM$.
GPU throughput requirements are not only significant, but vary by over $6\times$ across models.
The difference in throughput across the models is due to the variations in operational intensity (i.e., compute per sample) across models, as well as synchronization overheads between GPUs during each iteration.

Furthermore, we project the online preprocessing throughput requirement to increase by $3.5\times$ within the next two years due to larger training samples, improved hardware accelerators, and software optimizations.
We cannot simply over-provision resources for preprocessing at each trainer for the worst-case model; doing so would waste large amounts of capacity and power across our fleet.
The DSI pipeline must instead scale online preprocessing resources to meet intense and increasing GPU throughput demands, and adapt to the diverse requirements across models.

\subsection{Data Loading at GPU Trainers}\label{sec:data-loading}
Section~\ref{sec:dsi} described how our DSI pipeline allows training jobs to scale online preprocessing resources by loading preprocessed tensors from distributed \SystemName Workers.
We now show that data loading over the network at GPU trainers, even without extraction or transformations, still requires significant trainer CPU, network, and memory bandwidth.

\input{fig-frontend-utilization}
Figure~\ref{fig:frontend_utilization} shows the memory bandwidth and CPU utilization on the 2-socket, 8-GPU training node as we increase the rate at which preprocessed tensors are loaded from a set of \SystemName Workers.
Vertical lines represent the required GPU throughput across each $RM$, as measured in Table~\ref{tbl:gpu_throughput}.
High training data throughput demands driven by the GPUs directly translate to considerable front-end resource requirements for data loading.
Production-scale model training is approaching NIC saturation, even with significant reduction in data sizes due to transformation operations (see Section~\ref{sec:extract_transform}).
Second, even without expensive extraction or transformation operations, production models require up to $40\%$ of CPU cycles (almost a full socket) and $55\%$ of memory bandwidth to load training data.
This demand is due to network stack and memory management requirements in addition to the necessary "datacenter tax"~\cite{isca15_kanev} operations such as TLS decryption and Thrift deserialization that are required in our production environment.
Considering memory bandwidth saturates at $\approx70\%$ utilization, %
data loading constrains trainers' compute, memory bandwidth, and network resources. %

\subsection{Extracting \& Transforming Data at \SystemName Workers}\label{sec:extract_transform}
\input{tbl-reader-throughput}
Data extraction and transformation requires strikingly more resources than are available on trainers and must be distributed.

Table~\ref{tbl:reader_throughput} shows the maximum data extraction and transformation throughput achieved by each \SystemName Worker, running on a general purpose server (C-v1, Table~\ref{tbl:t1-skus}).
We need large and highly-variable throughput across models to meet the GPU demands in Table~\ref{tbl:gpu_throughput} --- between 9 and 55 servers per trainer node. %
Unlike traditional distributed query executors~\cite{sosp13-naiad, nsdi12-spark, osdi08-dryadlinq, vldb10-dremel}, which rely on large clusters to produce a result as fast as possible, online preprocessing requires continuous throughput guided by GPU demands; allocating more preprocessing workers will not improve end-to-end training time.
On the other hand, using trainer hosts for preprocessing is also insufficient, as our online preprocessing demands represent considerably more network, compute, and memory bandwidth resources than available locally, especially when factoring in data loading.
Not only do models require large and diverse data loading throughput, achievable extract and transform throughput, given fixed compute, varies across models,
emphasizing the need to right-size preprocessing resources to each model.

To understand the implications of these results in more detail, we observe that preprocessing significantly reduces data sizes, especially considering storage bytes are compressed in Table~\ref{tbl:reader_throughput}.
This is due to a combination of filtering, over reading features from storage (see Section~\ref{sec:future}), and size reduction during transformations.
This has implications on network throughput requirements, as $1.18$ to $3.64\times$ more network bandwidth is required to extract raw samples from storage than to load preprocessed tensors.
Thus, performing data extraction at the trainers would further amplify the network bandwidth requirements beyond the per-model requirements shown in Figure~\ref{fig:frontend_utilization}, resulting in data stalls.
Table~\ref{tbl:reader_throughput} shows how \SystemName Workers are bound on ingress NIC bandwidth for $RM_2$, requiring $\approx$ 10 Gbps of our current 12.5 Gbps NICs (Table~\ref{tbl:t1-skus}), reaching practical NIC throughput limits.
Even given higher NIC bandwidth limits, the datacenter network can pose potential bottlenecks, as upper datacenter network links are often oversubscribed~\cite{barroso_datacenter, sigcomm15-jupiter-rising, oic13-farrington, sigcomm15-roy}.

\input{fig-reader-utilization}
Furthermore, network bandwidth is not the only limiting factor.
Figure~\ref{fig:reader_utilization} shows \SystemName CPU and memory bandwidth utilization at saturation.
Each model exhibits diverse resource requirements.
$RM_1$ is bottlenecked on memory bandwidth and CPU utilization.
This is because $RM_1$ requires significantly more CPU cycles for preprocessing due to its computationally expensive transformations.
$RM_3$ is bound on memory capacity, forcing us to limit the worker thread pool size to avoid OOM exceptions.

\input{tbl-t1-skus}
While models are constrained on various hardware currently, these bottlenecks will change as future generations of compute nodes provide a different balance of compute, memory, and network.
Table~\ref{tbl:t1-skus} shows the hardware characteristics of our current (C-v1) and upcoming versions of compute nodes used by \SystemName.
To discount for the multifaceted and potentially non-technical factors impact the final specification of future compute nodes, we also show a hypothetical compute node (C-vSotA) based on state-of-the-art hardware (e.g., AMD Milan with 8 channels of DDR4-3200~\cite{amd-milan}, and a 100G NIC).
We observe an increased scaling in the number of cores and NIC bandwidth compared to memory bandwidth; we thus expect memory bandwidth to become the bottleneck as per-core memory bandwidth decreases compared to per-core NIC bandwidth.
To demonstrate this, we ran preprocessing for $RM_2$ on the C-v2 node in Table~\ref{tbl:t1-skus} and observed that memory bandwidth, not network, was the bottleneck.
We further studied where memory bandwidth is going, and observed that $50.4\%, 24.9\%, 16.4\%$, and $4.7\%$ of LLC misses were due to transformations, extraction, network receive, and network send, respectively, highlighting areas for future optimization.

\subsection{Transformation Operations}\label{sec:transformation-operations}
\input{tbl-gpu-preproc}
Accelerating transformation operations (e.g., via GPUs~\cite{dali}) are promising, but requires more research as our transformations can be distinctly different from the matrix-heavy operations used in training and may contend cycles with training.
Table~\ref{tbl:gpu-preproc} provides a list of important (but not exhaustive) preprocessing transformations that are needed by our production DLRMs.
These operations are distinctly different from the image-centric operations used in many preprocessing libraries~\cite{dali-ops}, such as crops, resizing, and color augmentations.

DLRM transformation operations can be split into three classes: feature generation, sparse feature normalization, and dense feature normalization.
Dense feature normalization (Logit, BoxCox, Onehot) and sparse feature normalization (SigridHash, FirstX) normalize features based on dataset statistics.
Feature generation operations (commonly-used ones include Bucketize, NGram, and MapId) derive new dense and sparse features from raw dataset features.
Feature generation is especially expensive --- dense normalization, sparse normalization, and feature generation typically require around $5\%$, $20\%$, and $75\%$ of transformation cycles, respectively.
We are actively open-sourcing these operations in TorchArrow~\cite{torcharrow}.

\subsection{Summary of Key Takeaways}
DLRM training jobs require online preprocessing that can induce data stalls on GPU trainers due to limited host compute, memory, and network resources.
We built \SystemName, a disaggregated online preprocessing service, to completely eliminate data stalls by offloading data extraction and transformation operations.
Nevertheless, trainers must be provisioned with enough host resources to handle data loading rates driven by the GPUs.
At disaggregated \SystemName Workers, we expect memory bandwidth to become the primary bottleneck, largely due to transformations.
Finally, we identified how DLRM transformation operations differ from image models, and feature generation dominates transformation compute. 

%% file: tbl-onbox.tex
\begin{table}[t]
  \centering
  \footnotesize
    \caption{\small Insufficient GPU trainer host resources for preprocessing introduces significant data stalls for $RM_1$.}%
  \label{tbl:onbox}
    \vspace{-3mm}
  \begin{tabular}{@{}rrr@{}}
    \toprule
    \% of GPU Stall Time & \% CPU Utilization & \% Memory BW Utilization \\ \midrule
    56                       & 92                & 54                      \\ \bottomrule
  \end{tabular}%
  \vspace{-3mm}
\end{table}

%% file: tbl-gpu-throughput.tex
\begin{table}[t]
  \centering
  \footnotesize
    \caption{\small RMs drive large and diverse throughput at each GPU training node.}
    \label{tbl:gpu_throughput}
    \vspace{-3mm}
    \begin{tabular}{@{}llll@{}}
      \toprule
                             & RM1  & RM2  & RM3 \\ \midrule
      GPU Trainer Throughput (GB/s, per 8-GPU Node) & 16.50 & 4.69 & 12.00  \\ \bottomrule
    \end{tabular}%
    \vspace{-3mm}
\end{table}

%% file: fig-frontend-utilization.tex
\begin{figure}[t]
  \centering
  \includegraphics[width=3.33in]{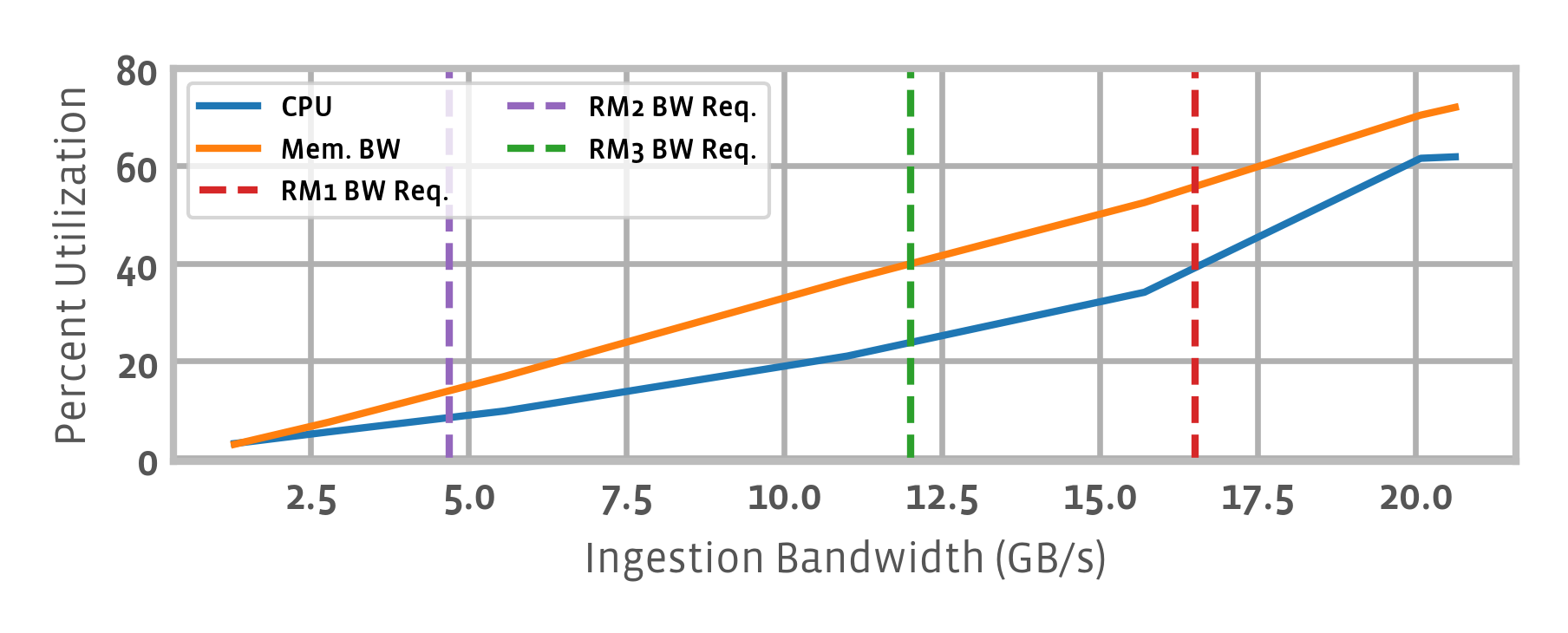} 
  \vspace{-8mm}
  \caption{\small CPU and memory bandwidth utilization at trainer frontend as data loading throughput scales. Vertical lines show the network utilization of each RM.}
  \label{fig:frontend_utilization}
  \vspace{-4mm}
\end{figure}

%% file: tbl-reader-throughput.tex
\begin{table}[t]
  \centering
  \footnotesize
  \caption{\small DPP Worker throughput across RMs and the resulting \# Workers required to meet trainer demands. Storage RX is compressed and transform RX/TX is uncompressed.}
  \label{tbl:reader_throughput}
  \vspace{-3mm}
  \begin{tabular}{@{}lrrrrr@{}}
    \toprule
    Model & kQPS & \thead{Storage\\ RX (GB/s)} & \thead{Transform\\ RX (GB/s)} & \thead{Transform\\ TX (GB/s)} & \thead{\# \SystemName Workers Req. \\ per GPU Training Node} \\ \midrule
    RM1 & 11.623 & 0.8 & 1.37 & 0.68 & 24.16 \\
    RM2 & 7.995  & 1.2 & 0.96 & 0.50 & 9.44  \\
    RM3 & 36.921 & 0.8 & 1.01 & 0.22 & 55.22 \\ \bottomrule
  \end{tabular}%
  \vspace{-5mm}
\end{table} 

%% file: fig-reader-utilization.tex
\begin{figure}[t]
  \centering
  \includegraphics[width=3.33in]{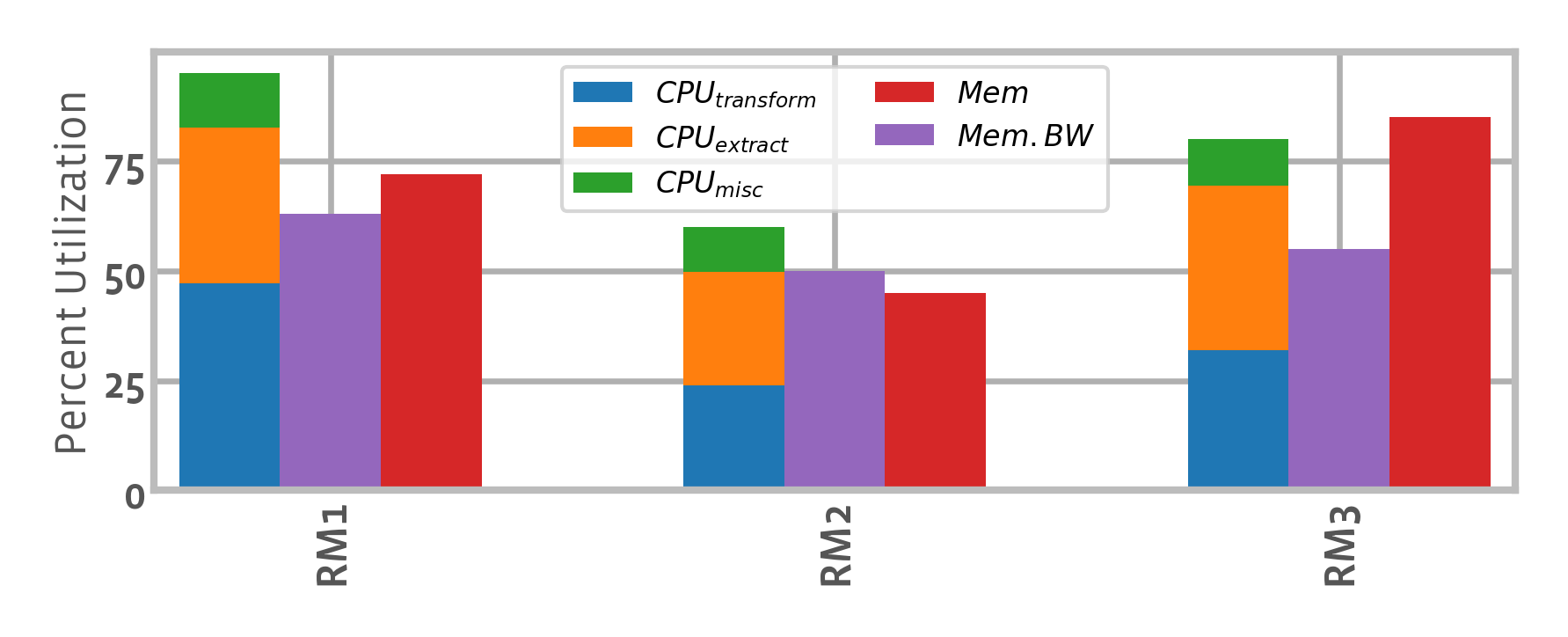}
  \vspace{-8mm}
  \caption{\small CPU, memory, and memory bandwidth utilization at preprocessing workers across RMs. CPU utilization is broken down into transformation, extraction, and miscellaneous cycles.}
  \label{fig:reader_utilization}
  \vspace{-3mm}
\end{figure}

%% file: tbl-t1-skus.tex
\begin{table}[t]
  \centering
  \footnotesize
  \caption{\small Hardware specifications across three versions of compute servers, plus a hypothetical state-of-the-art server.}
  \label{tbl:t1-skus}
  \vspace{-3mm}
\begin{tabular}{@{}lrrrrrr@{}}
\toprule
Node &  \thead{Num.\\ Physical\\ Cores} & \thead{NIC\\ (Gbps)} & \thead{Memory \\ (GB)} & \thead{Peak\\ Mem. BW\\ (GB/s)} & \thead{Peak Mem.\\ BW / Core\\ (GB/s)} & \thead{NIC BW\\ / Core\\ (Gbps)} \\ \midrule
C-v1 & 18       & 12.5       & 64          & 75     & 4.2 & 0.69 \\
C-v2 & 26       & 25.0         & 64          & 92     & 3.5 & 0.96 \\
C-v3 & 36       & 25.0         & 64          & 83     & 2.3 & 0.69  \\
C-vSotA & 64       & 100.0        & 1024     & 205 & 3.2  & 1.56  \\
\bottomrule
\end{tabular}%
\vspace{-4mm}
\end{table}

%% file: tbl-gpu-preproc.tex
\begin{table}[t]
\centering
\caption{\small Description of common preprocessing transformations.}
\label{tbl:gpu-preproc}
\vspace{-3mm}
\footnotesize
\begin{tabular}{@{}ll@{}}
\toprule
Op Name         & Description                                           \\ \midrule
Cartesian       & Compute Cartesian product between two sparse features \\
Bucketize       & Shard features based on bucket borders   \\
ComputeScore    & Arithmetic operations on sparse features              \\
Enumerate       & Similar to Python enumerate()                         \\
PositiveModulus & Compute positive modulus on sparse features           \\
IdListTransform & Performs intersection of two sparse feature lists     \\
BoxCox          & BoxCox transform for normalization                    \\
Logit           & Logit transform for normalization                \\
MapId           & Maps feature IDs to fixed values                      \\
FirstX          & List truncation of sparse features for normalization                   \\
GetLocalHour    & Compute local timestamp                               \\
SigridHash      & Compute hash value to normalize list of sparse features    \\
NGram           & Compute an n-gram between multiple sparse features     \\
Onehot          & Apply one hot encoding to normalize dense features                              \\
Clamp           & Same as std::clamp                                    \\
Sampling        & Randomly sample training data samples                       \\ \bottomrule
\end{tabular}%
\vspace{-5mm}
\end{table}

%% file: future.tex
\section{Key Insights and Research Opportunities}\label{sec:future}
This section assimilates and explores key insights we learned while architecting and profiling Meta's DSI pipeline and important research challenges we continue to face.
As DSAs for ML continue to improve, the DSI pipeline is becoming an increasingly resource-intensive component of the end-to-end ML training pipeline.
We argue that similar attention to DSI is warranted in order to continue scaling datacenter-scale training.

\subsection{Hardware Bottlenecks in DSI}\label{sec:future-bottlenecks}
\textbf{Storage layer.}
Cross-datacenter bandwidth is constrained (Section~\ref{sec:global-training}).
Thus, we must provision sufficient storage capacity and IOPS bandwidth within our production storage layer (i.e., Tectonic~\cite{fast21-tectonic}) in each datacenter. 
Storage capacity requirements are driven by the industry-scale dataset sizes (Table~\ref{tbl:table_size}). 
IOPS bandwidth requirements are driven by the overall trainer node throughput (Table~\ref{tbl:gpu_throughput}) and scaled by data volume changes due to preprocessing.
We must also factor in how I/O size characteristics (Table~\ref{tbl:io-size}) affect achievable IOPS of the HDD storage nodes.
As it stands, we observe an over 8$\times$ throughput-to-storage gap even after accounting for triplicate replication for durability.
In order to meet IOPS demands, we must provision significantly more storage capacity per datacenter than is required to store datasets, motivating the need for storage hardware and systems that better balance IOPS and storage capacity.

\textbf{Data ingestion layer.}
Section~\ref{sec:extract_transform} highlighted how the diverse set of preprocessing requirements across RMs constrained compute, network, and memory resources at \SystemName Workers.
However, Table~\ref{tbl:t1-skus} shows the expected growth of network bandwidth and compute capacity will outpace memory bandwidth for the next generations of our general-purpose compute nodes.
We expect data ingestion to be heavily constrained by memory bandwidth, warranting further research in methods to reduce memory bandwidth demand during preprocessing.

\textbf{Trainers.}
The goal of the DSI pipeline is to avoid data stalls, ensuring that trainers (and specifically DSAs) are fully utilized and dictate the throughput demands of the end-to-end training pipeline.
Section~\ref{sec:preprocessing} explains how without \SystemName, trainers are constrained on both front-end network and host memory and compute resources due to online preprocessing.
After disaggregating \SystemName, we can achieve our goal of ensuring that the accelerators are fully fed by simply provisioning enough host compute, memory bandwidth, network resources for data loading when designing our training nodes.
For example, our next-generation ZionEX nodes contain 4 CPU sockets, each with a 100 Gbps dedicated front-end NIC, to ensure that all GPUs are fully fed~\cite{arxiv-mudigere}.

\subsection{Heterogeneous Hardware for DSI}\label{sec:future-heterogeneous}
The bottlenecks above allude to ample opportunity to leverage heterogeneous hardware across data storage and ingestion. %

\textbf{Balancing storage throughput and capacity.}
We noted how our HDD-based storage nodes presented $\approx8\times$ throughput-to-storage gap, requiring us to provision datacenters with excess storage capacity to meet IOPS demand.
We can improve the power efficiency of our storage layer by balancing storage throughput and capacity by leveraging heterogeneous storage media with a higher IOPS per watt.
For example, our SSD-based storage nodes can provide $326\%$ IOPS per watt, but trades off storage capacity with only $9\%$ capacity per watt, compared to HDDs within our fleet.
At the same time, simply placing training data on SSDs would result in an unfavorable storage-to-throughput gap due to our large datasets.

Storage layers for DSI should balance storage throughput and capacity for optimal power efficiency.
There are further software and hardware optimization opportunities, such as placing commonly-used features (Figure~\ref{fig:reuse_analysis}) on SSD-based caches, or leveraging non-volatile memory.
These solutions must consider important characteristics of industry-scale DSI pipelines characterized in Sections~\ref{sec:training} and \ref{sec:dataset}.
Datacenter architects must balance storage and IOPS capacity in each datacenter while reasoning about highly-variable training demand, shared capacity across multiple datasets, and dynamic scheduling behavior across geo-distributed datacenters.
The storage layer must also accurately predict and place commonly-used bytes on the appropriate medium, requiring complex predictions on locality of highly asynchronous training jobs and the reuse patterns of continuously-evolving feature sets and model architectures.

\textbf{Accelerating transformations.}
We also foresee further opportunities for acceleration during online preprocessing.
For example, recent efforts have been used to accelerate preprocessing on the training GPU~\cite{dali}, but risks degrading training throughput.
While we believe preprocessing can be accelerated, there are numerous challenges to address.

Transformations can be performed on the GPU trainer, host CPU, disaggregated CPUs, or disaggregated accelerators --- deciding the optimal placement is non-trivial.
Many transformation operations described in Section~\ref{sec:preprocessing} exhibit diverse amenability for acceleration; the most efficient hardware platform can vary across operations.
For example, we observed an $11.9\times$ and $1.3\times$ GPU/CPU performance for SigridHash and Bucketize, respectively, on a V100 GPU and 20 CPU threads.
The prevalence of each operation, and thus the most efficient preprocessing solution, also varies heavily across models.

In fact, each model requires a large graph of many operations across its feature set.
For example, a single feature $X$ may require a DAG of multiple operations that apply \textit{Bucketize} to feature $A$, apply \textit{FirstX} to feature $B$, compute the \textit{Ngram} of the intermediate values, and apply \textit{SigridHash} to generate feature $X$.
Current GPU hardware and APIs are optimized for large, parallel singular tasks as opposed to multiple small, diverse operations. 
Launching a kernel for each feature incurs significant launch and host-to-device transfer overheads.
For example, we observed over three orders of magnitude throughput speedup by applying a simple kernel on a tensor combining ~1000 sparse features versus applying the same kernel on each feature separately.
Furthermore, transformation operations produce outputs and intermediate values with variable-length sequences, requiring complex memory management not well addressed by current GPUs and ML frameworks~\cite{ppopp21-turbotransformers}.
Addressing these characteristics of preprocessing is critical in order to fully leverage the acceleration potential of GPUs, FPGAs, or other DSAs.

\textbf{Accelerating data extraction and loading.}
Accelerators can be applied beyond transformation operations.
Section~\ref{sec:preprocessing} showed how datacenter tax operations further constrain \SystemName Worker resources.
Necessary TLS operations amplify memory bandwidth by $3\times$, further constraining the limited bandwidth resource.
Hardware such as accelerators for microservices~\cite{asplos20-accelerometer} and SmartNICs that support techniques such as TLS offloading will be needed to further scale DSI.

\subsection{Datacenter Planning and Scheduling}\label{sec:future-datacenter}
We must intelligently design and provision compute, network, and storage capacity in each datacenter to ensure both high utilization of each DSI system given a fixed power budget and sufficient capacity to meet peak training demands.
To do so, we rely on extensive models built by continuously profiling DSI workloads on both storage nodes and \SystemName Workers.
Designing a datacenter for ML training thus requires not only understanding compute requirements of the models, but also an accurate benchmark of the datasets and preprocessing requirements (which we discuss next).

We must also consider scheduling as an important component of ML training systems.
We have multiple datacenters in a region and multiple regions globally.
Section~\ref{sec:training} explored how our training workload is spread across regions, requiring datasets to be replicated and co-located with trainers.
We foresee opportunity for a global scheduler to intelligently route and bin-pack training jobs to specific regions to reduce storage duplication.
Furthermore, as our datasets continue to grow beyond DC-scale, other parallelism techniques, such as model-hopper parallelism~\cite{vldb20-cerebro} to move TB-scale models across DCs instead of PB-scale data, will become increasingly important.

\subsection{Representative Benchmark Datasets}\label{sec:future-benchmarks}
Recent advancements in DSAs have largely been driven and measured by benchmarks.
For example, the latest round of MLPerf Training~\cite{mlsys20-mlperf} (v1.1) saw results from 14 organizations with up to 2.3$\times$ improvement over the previous round.
Unfortunately, these benchmarks have largely focused on models as opposed to datasets, leading to rapid innovation in model architectures and training nodes while largely ignoring the DSI pipeline.
Current leading benchmark datasets, such as ImageNet~\cite{imagenet} and COCO~\cite{coco}, were released in 2010 and 2014 and have been largely unchanged, despite their ubiquity in academia.
The Criteo 1TB Click Logs~\cite{criteo}, the Recommendation dataset used by MLPerf Training~\cite{wu:2020:mlperfrec}, was released in 2013. 

Section~\ref{sec:dataset} describes how production recommendation datasets differ from static benchmark datasets.
Understanding how representative datasets are generated, stored, and read are critical to further research in DSI for ML training.
Specifically, we characterized that for DLRMs, training samples are a) continuously generated from real-world events, b) stored as structured samples in a common data warehouse or feature store which are further stored as columnar files in a distributed filesystem, and c) require petabytes of storage.
Furthermore, when training jobs read the dataset, they d) perform extensive online preprocessing operations, e) only require one epoch, and f) require further row-wise and feature-wise filtering.
We are actively working towards more representative benchmark datasets (e.g., \cite{dlrm-dataset}), and we hope this paper motivates the importance of further work in this area.

\subsection{Multi-dimensional System Co-design}\label{sec:future-optimization}
Efficiency gains solely via hardware are slow, as hardware refreshes must be planned years in advance.
To continuously improve DSI efficiency (and increase training capacity), \textit{we spend significant effort optimizing our DSI systems because at scale, small efficiency gains can translate to MWs of additional trainer capacity}.
In our experience, DSI system architects must understand and co-design optimizations across two dimensions.
\textbf{Top-to-bottom:} DSI systems must leverage characteristics and meet requirements of applications (ML models and engineers) while optimizing for the underlying hardware.
\textbf{End-to-end:} Optimizations must be considered across the DSI components; optimizations can trade-off efficiency across the entire pipeline.
We highlight this through recent examples next.

\input{fig-flatten-regroup}
\input{tbl-aggregate}

\textbf{Feature flattening.}
Our warehouse tables' schema, which store features in maps, are designed to handle constantly-evolving feature sets.
This required training jobs to read the entire row, resulting in a large "over read" of bytes from storage as each training job only requires a small set of features from each row.
To maintain the usability benefits (i.e., rapid feature engineering) of our map schema for applications while avoiding over reads, we optimized our DWRF file format with \textit{feature flattening}.
As shown in Figure~\ref{fig:flatten_regroup_example}, feature flattening organizes maps such that values for a given map key (feature ID), across rows, are stored as separate streams on disk.
Combined with additional metadata for each feature ID in the DWRF file, readers can selectively read the desired features from storage.
Table~\ref{tbl:aggregate} shows how feature flattening doubled \SystemName Worker throughput (with $12\%$ increase in storage capacity) due to a reduction in CPU cycles spent extracting unnecessary data.

\textbf{Coalesced reads.}
Unfortunately, filtering at the storage layer reduced I/O sizes from almost 8 MB (Tectonic's chunk size) to the small I/O sizes ($\approx 20$ KB) shown in Table~\ref{tbl:io-size}, resulting in poor IOPS on our HDD-based storage nodes due to excessive disk seeks.
This reduced storage throughput by $97\%$, as shown in Table~\ref{tbl:aggregate}.
We optimized top-to-bottom for our HDDs by \textit{coalescing reads}.
Figure~\ref{fig:flatten_regroup_example} shows how coalesced reads group selected feature streams within $1.25$ MiB in one I/O, eliminating storage throughput degradation by amortizing disk seeks.

\textbf{Feature reordering.}
Coalesced reads resulted in over reads of unused features, limiting its effectiveness. 
These over reads occur because the offline data generation step effectively orders feature streams randomly.
In the example in Figure~\ref{fig:flatten_regroup_example}, reading features (A, D) with a coalesced read ends up over reading (B, C), squeezed between (A, D).
We applied end-to-end optimization by augmenting our data generation path to continuously write feature streams in each file ordered based on features' popularity in training jobs launched within a recent window (e.g., 7 days).
Feature reordering leverages the insight that a small set of popular features contribute to a large percent of storage throughput (Section~\ref{sec:reuse-dataset}), reducing the amount of unnecessary features in a coalesced read and improving storage throughput by $84\%$.

\textbf{Large stripes, in-memory flatmaps, and localized optimizations.}
Finally, we applied numerous other optimizations across the DSI pipeline.
We used \textit{large stripes} that increase the number of rows in each file stripe to increase the average I/O size of each read.
Increasing stripes to $\approx$ 1 GB further improved storage throughput by $31\%$.
We also noted how both DWRF and tensor formats represent each feature's values contiguously across rows, while data extraction reconstructs all features in a row-based map format, requiring costly format changes and copies between columnar and row formats during preprocessing.
We changed how DPP Workers represent samples with \textit{in-memory flatmaps} to match the DWRF and tensor formats.
This reduced format conversions and thus reduced constrained memory bandwidth demands, improving Worker throughput $15\%$.
Finally, even localized optimizations at \SystemName Workers, such as removing unnecessary null checks and using build-time optimizations like Linker Time Optimization (LTO) and AutoFDO~\cite{cgo16-autofdo}, further improved \SystemName throughput by $28\%$.

The optimizations above required us to consider DSI optimizations top-to-bottom and end-to-end.
For example, combining feature flattening, coalesced reads, and feature reordering considered ML application characteristics and demands, such as filtering on evolving feature sets, while addressing seek overheads of our underlying HDD hardware (top-to-bottom optimization).
Similarly, we demonstrated how feature flattening trades-off storage capacity for \SystemName efficiency, as well as how optimizations such as large stripes, feature reordering, and in-memory flatmaps are made across the end-to-end DSI pipeline from dataset generation to online preprocessing.

In total, Table~\ref{tbl:aggregate} shows how these optimizations increased \SystemName and storage throughput by $2.94\times$ and $2.41\times$, respectively.
When weighed by our provisioned DPP and storage power requirements, these co-designed optimizations resulted in a $2.59\times$ reduction in DSI power requirements, allowing us to provision datacenters with significantly more compute resources for trainers.

We are continuing to explore promising co-designed optimization opportunities that aim at improving data extraction and optimizing \SystemName in-memory formats, such as Velox~\cite{velox} and TorchArrow~\cite{torcharrow}.
We are also exploring other optimization techniques, such as caching preprocessed tensors and balancing transformations between offline and online ETL.

%% file: fig-flatten-regroup.tex
\begin{figure}[t]
  \centering
  \includegraphics[width=3.33in]{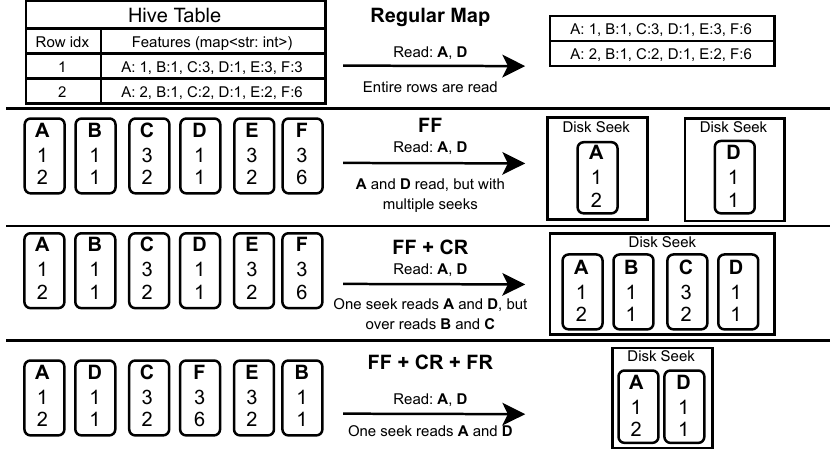}
  \caption{\small Example highlighting which features are read given regular map, \textbf{FF}=feature flattening, \textbf{CR}=coalesced reads, and \textbf{FR}=feature reordering.}
  \vspace{-3mm}
  \label{fig:flatten_regroup_example}
\end{figure}

%% file: tbl-aggregate.tex
\begin{table}[t]
\footnotesize
\caption{\small Normalized \SystemName Worker and storage throughput as a result of progressive optimizations. FF = Feature Flattening, FM = In-Memory Flatmap, LO = Localized Optimizations, CR = Coalesced Reads, FR = Feature Reordering, LS = Large Stripes.}
\label{tbl:aggregate}
\vspace{-3mm}
\begin{tabular}{@{}lrrrrrrr@{}}
\toprule
                   & Baseline & +FF & +FM & +LO  & +CR   & +FR   & +LS   \\ \midrule
\SystemName Throughput  & 1.00        & 2.00    & 2.30  & 2.94 & 2.94  & 2.94 & 2.94  \\
Storage Throughput & 1.00        & 0.03 & 0.03 & 0.03 & 0.99  & 1.84 & 2.41 \\ \bottomrule
\end{tabular}%
\end{table}

%% file: related.tex
\section{Related Work}\label{sec:related}
\noindent\textbf{Data storage and ingestion for ML.}
We presented Meta's production deployed DSI pipeline.

\textit{\underline{ETL pipelines.}}
We discussed how we use traditional ETL engines, such as Spark~\cite{nsdi12-spark}, to generate structured training data from raw logs.
A number of query and streaming engines~\cite{sosp13-naiad, nsdi12-spark, vldb10-dremel, osdi08-dryadlinq, sosp13-sparkstreaming, vldb15-beam} are used across industry for this task.
We characterized how online data preprocessing demands diverge from traditional ETL, requiring deep integration into PyTorch, pipelined compute, localized mini-batch transforms, and right-sizing, highlighting a need for a distinct online preprocessing framework that optimizes for power efficiency while eliminating data stalls.

\textit{\underline{Data storage and warehousing for ML.}}
We characterized how industrial datasets differ from benchmarks, requiring massive and evolving datasets, highly selective filtering, and interoperability and reuse across multiple models and systems.
To address these needs, we store datasets as Hive~\cite{vldb09-hive} tables on top of Tectonic~\cite{fast21-tectonic} using an optimized Apache ORC~\cite{apache-orc} like format.
Comparable solutions exist across industry.
Feature stores (e.g., Tecton~\cite{tecton}) and data warehouses (e.g., DeltaLake~\cite{vldb20-deltalake} and Snowflake~\cite{sigmod16-snowflake}) manage datasets.
These rely on variety of storage and memory formats~\cite{apache-avro, apache-parquet, apache-arrow, tfrecord}.

\textit{\underline{Online preprocessing for ML.}}
tf.data~\cite{tfdata} presents a runtime and API for online preprocessing in TensorFlow~\cite{osdi16-tensorflow}.
While tf.data Service~\cite{tfdata-service} is an experimental feature to distribute online preprocessing on a user-managed cluster, tf.data focuses on optimizing preprocessing on the host CPU.
\SystemName similarly presents a runtime for online preprocessing fully integrated in PyTorch~\cite{neurips19-pytorch}.
\SystemName is inherently disaggregated and runs as a fully-managed service at Meta, enabling online preprocessing to automatically scale to meet the throughput required by training jobs running on hundreds of GPUs.

Other recent works target key DSI components, focusing on benchmark vision and NLP models.
CoorDL~\cite{vldb21-mohan}, Quiver~\cite{fast20-quiver}, and DIESEL~\cite{icpp20-diesel} are caches that optimize for single-server training, HP tuning jobs, and small files, respectively.
DeepIO~\cite{mascots18-deepio} and DLFS~\cite{cluster19-dlfs} leverage hardware (RDMA and NVMeOF) to provide randomized minibatches from storage.
Revamper~\cite{atc21-revamper} randomly augments samples across epochs to reduce online preprocessing costs.
Wang \textit{et al.}~\cite{mlsys20-wang} and Kumar \textit{et al.}~\cite{kumar21exploring} mitigate data stalls on TPUs.
OneAccess~\cite{hotcloud19-oneaccess} motivates sharing online preprocessing for HP tuning jobs.
Industrial DSI characteristics are markedly different from benchmarks (Section~\ref{sec:future}), limiting the impact of such systems in industrial settings. %

\noindent\textbf{Understanding Large-Scale Training.}
tf.data~\cite{tfdata} characterized online preprocessing at Google, highlighting similar findings such as prevalent data reuse and demanding compute requirements.
Xin \textit{et al.}~\cite{sigmod21-xin} characterized how ML models are trained and deployed at Google, focusing on model lifecycle management.
Pulkit \textit{et al.}~\cite{sigmod19-mldp} presented MLdp, a data management platform at Apple, focusing on industrial data lifecycle management requirements including versioning, provenance, and access control.
Hazelwood \textit{et al.}~\cite{hpca18-hazelwood} presented how Meta trains and serves ML models at scale.
To the best of our knowledge, our work represents the first characterization of the end-to-end DSI workloads, systems, and infrastructure in a large-scale training deployment, noting key research opportunities.

\noindent\textbf{Recommendation Models.}
Gupta \textit{et al.}~\cite{hpca20-gupta} analyzed industry-scale inference models on three different CPU architectures.
DeepRecSys~\cite{isca20-deeprecsys} and RecSSD~\cite{asplos21-recssd} optimized inference requests across CPUs and GPUs, and SSDs, respectively.
Acun \textit{et al.}~\cite{hpca21-acun} characterized DLRM architectures on GPU trainers, 
Sethi \textit{et al.}~\cite{asplos22-sethi} presented an optimized embedding sharding strategy for DLRM training,
Maeng \textit{et al.}~\cite{mlsys21-cpr} explored checkpointing trainer state, and AIBox~\cite{cikm19-aibox} optimized training using hierarchical memory for parameters.
However, these prior works do not discuss the DSI pipeline, a critical part of ML training.

%% file: conclusion.tex
\section{Conclusion}\label{sec:conclusion}
DSI infrastructure will dominate large-scale training resource and power capacity without further innovation and optimization.
This paper presented Meta's end-to-end data storage, ingestion, and training pipeline used to train our production recommendation models.
We characterized DSI workloads on our fleet, including coordinated training, dataset storage and reading, and online preprocessing workloads.
To spur further research, we synthesized key insights and important research directions gleaned from our characterization and experience.